\documentclass[11pt,a4paper]{article}
\usepackage{jheppub}
\usepackage{xcolor}
\usepackage{amsmath,amssymb,amsfonts}
\usepackage{physics}
\usepackage{bbm}
\usepackage{graphicx}
\usepackage{svg}
\graphicspath{{./images}}
\usepackage{subcaption}
\usepackage{array}
\usepackage{tikz}
\usetikzlibrary{arrows.meta, positioning}
\usepackage{url}
\usepackage{multirow}
\usepackage{tabularray}
\usepackage{adjustbox}
\usepackage{comment}
\usepackage{tabularx}
\usepackage{slashed}
\newcolumntype{Y}{>{\centering\arraybackslash}X}
\usepackage{makecell}
\newcommand{\eq}[1]{eq.~(\ref{#1})}

\newcommand{\bea}{\begin{eqnarray}}
\newcommand{\eea}{\end{eqnarray}}

\newcommand{\nn}{\nonumber}

\usepackage{hyperref}

\usepackage{hyperref}

\begin{document}

\definecolor{lime}{HTML}{A6CE39}
\DeclareRobustCommand{\orcidicon}{\hspace{-1mm}
	\begin{tikzpicture}
	\draw[lime, fill=lime] (0,0) 
	circle [radius=0.12] 
	node[white] {{\fontfamily{qag}\selectfont \tiny \,ID}};
	\draw[white, fill=white] (-0.0525,0.095) 
	circle [radius=0.007];
	\end{tikzpicture}
	\hspace{-3mm}
}

\foreach \x in {A, ..., Z}{\expandafter\xdef\csname orcid\x\endcsname{\noexpand\href{https://orcid.org/\csname orcidauthor\x\endcsname}
		{\noexpand\orcidicon}}
}
\newcommand{\orcidauthorA}{0009-0009-3206-9970}
\newcommand{\orcidauthorB}{0009-0000-2346-2273}
\newcommand{\orcidauthorC}{0000-0003-4398-4698}

\title{A Lighter  QCD-Axion from Small Instantons}
\preprint{TIFR/TH/26-32} 

\author[a]{Sourav Bera\orcidA{}}
\author[b]{Susobhan Chattopadhyay\orcidB{}}
\author[a]{Rick S. Gupta\orcidC{}}
\affiliation[a]{Department of Theoretical Physics, Tata Institute of Fundamental Research, Homi Bhabha Road, Colaba, Mumbai 400005, India}
\affiliation[b]{Department of Physics and Astronomy, University of Utah, Salt Lake City, UT, 84112, USA}

\emailAdd{sourav.bera@tifr.res.in}
\emailAdd{susobhan.chattopadhyay@utah.edu}
\emailAdd{rsgupta@theory.tifr.res.in}

\begin{abstract}
{We propose models where the QCD axion mass can be much smaller than the standard expectation. We show that this  is possible in  models where both  QCD and the Standard Model flavor structure get modified in the ultraviolet (UV). In these models, both the axion mass and the up-quark Yukawa can receive sizeable  non-perturbative   contributions from instantons in the UV. If both these effects are carefully taken into account, the QCD axion is not necessarily heavier than usual expectations  and can, in fact, be much lighter. We show that if  small instanton contributions  account for a large fraction of the up-quark Yukawa, the QCD axion  can be  lighter than standard expectations whereas it is heavier otherwise. We propose two different models where this condition is realised: (1) one where the QCD gauge group emerges as a subgroup from a product of $SU(3)$ groups in the UV; and  (2) a composite Higgs model that also implements partial compositeness.  In both these models  the QCD axion can lie not only in the region to the right of the QCD axion band but also far to its left. This is especially relevant for a range of proposed haloscope and helioscope experiments that will probe a region mostly  to the left of the QCD band. 

}
\end{abstract}

\maketitle
\setcounter{page}{0}
\pagenumbering{arabic}

\section{Introduction}

The QCD axion~\cite{Peccei:1977hh,Peccei:1977ur,Weinberg:1977ma,Wilczek:1977pj,Kim:1979if,Shifman:1979if,Dine:1981rt,Zhitnitsky:1980tq} is one of the most promising candidates for new physics beyond the standard model. Not only does it provide an elegant solution to the strong CP problem, it is also one of the leading dark matter candidates~\cite{Preskill:1982cy,Abbott:1982af,Dine:1982ah}.  From an experimental point of view, one of the most  attractive features of the QCD axion  is that using only the low energy measurements of the pion mass and
decay constant, the axion mass and its coupling can be predicted to lie in a narrow band in the parameter space, the so-called `QCD axion band'. This has led to a blooming of  experimental ideas---including many proposed in the last decade---that aim to probe almost the whole QCD axion band in the coming years.

While this QCD axion band provides a clear-cut target for experimentalists, the prediction that the QCD axion must lie inside it rests on the assumption that  QCD itself  is not modified in the ultraviolet (UV). With this assumption, the axion mass and---within ${\cal O}(1)$ factors--- its couplings can be predicted in a robust way just from the knowledge of its decay constant, $f_a$. This is because  the only source of the breaking of the axion shift symmetry   is  non-perturbative QCD  in the infrared (IR).  The QCD axion mass can thus be reliably calculated using  the chiral lagrangian. While it is straightforward to introduce new sources of shift symmetry breaking, these new corrections are in general not aligned in phase with the QCD contribution and thus typically spoil the axion solution to the strong CP problem.

On the other hand, if QCD does get modified at high energies---for instance if the gauge coupling becomes large  again in the UV---it can provide a new aligned contribution to the axion mass from small instantons at short distance scales. This provides one of the principal ways to evade the prediction that the  axion must lie within the QCD axion band. This avenue  has been explored  in detail in the literature~\cite{HoldomPeskin1982,Holdom1985,FlynnRandall1987,
Gaillard:color_unified_axion,Agrawal:many_heavy_axion,
Gherghetta:5D_small_instantons,Csaki:many_heavy_axion,
Gherghetta:composite_Higgs_axion,Kitano:small_instantons,
RSG_Comp_Higgs,Kivel:supersizing_axions,Csaki:Instanton_NDA}. Other mechanisms that allow QCD axions to lie outside the QCD band include modified  couplings~\cite{DiLuzio:2016sbl,Farina:photophilic,Agrawal:2017cmd,
MarquesTavares:hadronic_axion,DiLuzio:astrophobic_axions,
Bjorkeroth:electron_decoupling,Darme:selective_enhancement} using the clockwork mechanism~\cite{Choi:2014rja,Choi:2015fiu,Kaplan:2015fuy} or by other means;
contributions from additional confining sectors~\cite{
Rubakov:heavy_axion,Fukuda:visible_axion,
Hook:discrete_hierarchy,DiLuzio:lighter_axion};
and Peccei--Quinn-preserving mass mixing with other
pseudoscalars~\cite{Gavela:axion_sum_rule}.

All the models involving small instantons above discuss the possibility of a QCD axion heavier than the standard value. This might seem natural because small instantons provide a new contribution to the axion mass in addition to the IR one from the chiral lagrangian. Even the other mechanisms listed above typically increase  the axion mass. Indeed there is only one mechanism in the literature that achieves a  QCD axion that is naturally lighter than standard expectations for a given $f_a$---namely the $\mathbb{Z}_N$-axion of Ref.~\cite{Hook:discrete_hierarchy, DiLuzio:lighter_axion, banerjeehook}.\footnote{See Ref.~\cite{Elahi:2023anarchy,Co:2024hilltop} for models having a light QCD axion that involve fine-tuning.}  This mechanism involves an axion potential that gets contributions from multiple confining sectors that cancel each other  giving only a small residual mass; the cancellation is the result of   a discrete $\mathbb{Z}_N$ symmetry that relates the different sectors.  While this is an ingenious mechanism it is somewhat contrived as the only purpose served by the multiple sectors is to engineer  a setup where contributions to the QCD axion   cancel among themselves. Moreover, in its simplest form, the model has a vacuum selection problem that leads to an ${\cal O}(1/N)$ tuning (see however Ref.~\cite{banerjeehook}).

In this work we show that the conventional wisdom that small instantons can only increase the axion mass rests on  the assumption that the SM flavor sector is unaltered in the UV. We show that if the SM flavor sector is modified in the UV, small instantons can not only increase the axion mass but also decrease it relative to the  usual expectation. This is because small instantons not only contribute to the axion mass but also give contributions that can account for a part of the up quark mass in the IR. We show  that if these small instanton contributions account for a large fraction of the up mass, the QCD axion, can in fact, be much lighter than standard expectations. If one assumes no modification of the SM flavor sector, however, these contributions to the up mass are proportional to the product of all the other SM Yukawa couplings resulting in a huge suppression of this effect. Thus the two main prerequisites to obtain a light QCD axion  are a UV modification of QCD and a UV modification of the SM flavor sector---the latter being something we might, in any case, expect   given the SM flavor puzzle.

In this work we propose two QCD axion models where small instantons not only give a contribution to the axion mass but also account for a large fraction of the up quark mass in the IR. As a result the QCD axion can be lighter than the usual expectation in both these models. The first model is a variant of the model in Ref.~\cite{Agrawal:massless_up} where QCD arises from a product of $SU(3)$ gauge groups in the UV. The individual gauge groups can have a much larger gauge coupling than QCD, making small instantons unsuppressed in the UV. The flavor suppression can be overcome in these models because not all SM quarks are charged under the same $SU(3)$ group. The second model we propose is based on the composite Higgs model in Ref.~\cite{RSG_Comp_Higgs} that also has a QCD axion. Due to the  presence of the large number of  colored constituent fermions, the QCD coupling increases in the UV in this model thus enhancing small instanton contributions. This model realises partial compositeness where the SM mass and mixing hierarchies arise in the IR from an anarchic  flavor structure in the UV; this allows the model to overcome the SM flavor suppression in the small instanton contributions discussed above. While we  discuss  only two models in detail, our arguments may apply to other scenarios where the QCD and flavor suppressions to small instanton contributions can be overcome; one such candidate is the set-up described in Ref.~\cite{koren}.

The QCD axion in these models can inhabit the region both to the left and right  of the QCD band. As we discuss in detail in Sec.~\ref{pheno}, this---in particular the fact  that the QCD axion can also lie  to the left of the band---strengthens the case for multiple proposed experiments that will probe precisely  this region of the parameter space.

\section{The QCD axion mass in the presence of small instantons}
\label{sec2}
In this section we discuss how the standard computation of the QCD axion potential is altered in the presence of small instantons. To this end, we present a toy model in Sec.~\ref{toy} where   QCD instantons become unsuppressed at some high energy scale, $M$,  and then derive the corresponding low energy  lagrangian below this scale.  The form of the lagrangian thus obtained is, in fact,  more general than the specific UV toy model; in particular, it captures all the effects arising from the more realistic models  presented in Sec.~\ref{product}. and Sec.~\ref{composite}.  We then derive the general lagrangian below $\Lambda_{\rm QCD}$ and show that the axion mass can be both lighter and heavier than standard expectations in the presence of small instantons.

\subsection{Toy model with unsuppressed small instantons and a QCD axion}
\label{toy}
\begin{figure}[t]
\label{toysi}
\begin{subfigure}{0.5\textwidth}
\hspace{-0.5cm}
\includegraphics[width=0.95\linewidth]{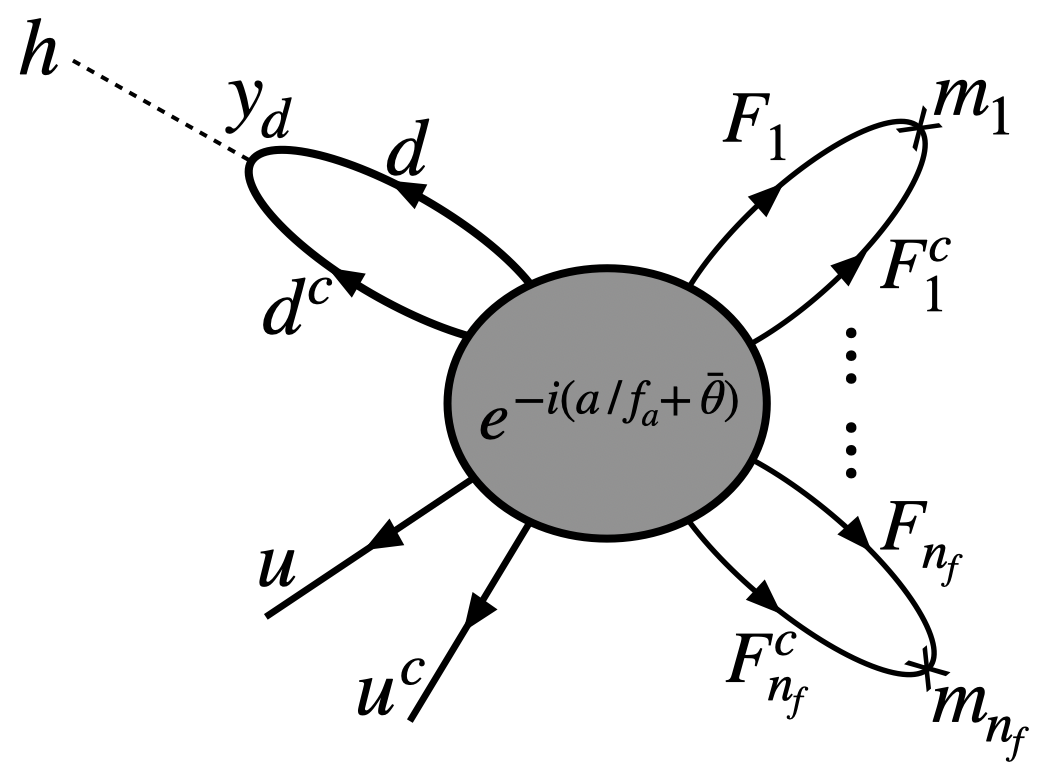}
\caption{}
\label{toysia}
\end{subfigure}
\begin{subfigure}{0.5\textwidth}
\includegraphics[width=0.9\linewidth]{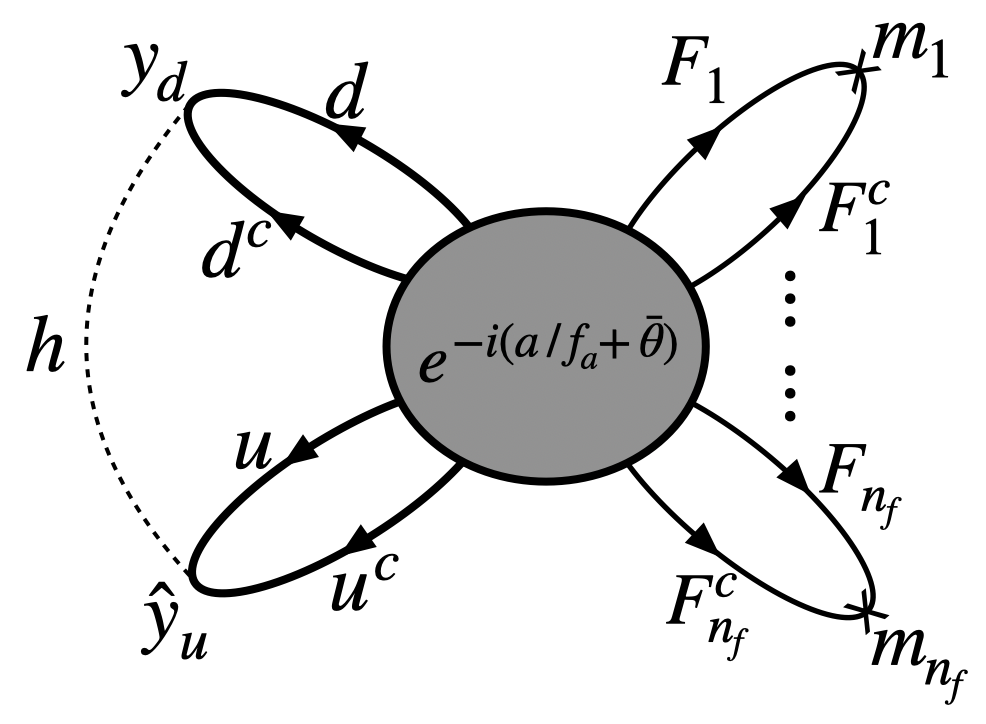} 
\caption{}
\label{toysib}
\end{subfigure}
\hspace{0.5cm}
\caption{Small instanton contributions to (a) the up-quark Yukawa, and (b) the axion potential  in the toy model of Sec.~\ref{toy}.  We do not show  here a  contribution---obtained by swapping the up and down quark labels in the left figure---that gives an additive correction to the down-quark Yukawa.}
\label{toysi}
\end{figure}

Consider a model that is exactly like the SM but with three important  differences: (1)  it contains a single generation of fermions whose Yukawa couplings differ from those of the corresponding SM fermions; (2) it contains an axion field, $a$; and (3) it contains some extra colored vector-like  fermions, $F_i$. The lagrangian  terms relevant to our discussion are as follows,
\bea
  {\cal L}\supset\frac{(\partial_\mu a)^2}{2}+
 \frac{g_s^2}{16\pi^2}\left(\frac{a}{f_a}+\bar\theta\right)
    \text{Tr}\,G_{\mu\nu}\widetilde G^{\,\mu\nu}-\left( \hat{y}_u \frac{h}{\sqrt{2}}u\,u^c + {y}_d  \frac{h}{\sqrt{2}}  d\,d^c+ \sum_{i=1}^{n_F} m_i F_{i}\,F_{i}^c + h.c \right),\nonumber\\
    \label{ltoy}
  \eea
where $h$ is the Higgs boson and the fermions, $u,u^c, d, d^c, F_i, F_i^c$ are all left-handed  Weyl spinors. In the above lagrangian we have  chosen the Yukawa couplings,  $\hat{y}_u$ and $y_d$, and the masses, $m_i$,    to be real as  any phase appearing in these terms can be rotated into the second term by a chiral rotation of the quark fields.

 We will consider a scenario where $\hat{y}_u \ll {y}_d$.  In the  limit,  $\hat{y}_u \to 0$, the  lagrangian in \eq{ltoy} has three  global symmetries at the classical level: (1) the symmetry, $u \to e^{i \alpha_1}u,\, u^c \to  e^{-i \alpha_1}u^c$; (2) the axial symmetry: $u \to e^{i \alpha_2}u,\, u^c \to  e^{i \alpha_2}u^c$; and (3) the Peccei-Quinn (PQ) symmetry, $a/f_a \to a/f_a +  \alpha_3$.  The second and third   symmetries above are, in fact, anomalous but their following  linear combination,  
 \bea
u \to e^{i\alpha} u ,\,~~~~  u^c \to e^{i\alpha} u^c ,\,~~~~ \frac{a}{f_a}\to \frac{a}{f_a}+2 \alpha,
\label{global1}
\eea
is anomaly free. The symmetry defined above  thus becomes  exact in the limit $\hat{y}_u \to 0$.
 
We assume that there are a sufficient number of vector-like fermions such that  the sign of the QCD beta function,
\begin{equation}
    \frac{d g_s}{d\ln\mu} = -\left(11 - \dfrac{2}{3}(n_F+2)\right) \frac{g^3_s}{16 \pi^2}
\end{equation}
is positive  at scales greater than the mass of the heaviest vector-like fermion, $\mu \gg m_i^{max}$. The  QCD coupling, $g_s$, then runs to a large value, $g_s(M)\gg 1$, at some higher scale, $M$. At scales greater than $M$  there must be some mechanism to tame the growth of $g_s$; we keep this physics  unspecified in this toy model.

At the scale, $\mu \sim M$, the QCD instantons become   unsuppressed giving rise to  the   two contributions shown in Fig.~\ref{toysi}. The first is a contribution to the up Yukawa and the second to the axion potential. The first contribution in Fig.~\ref{toysia}  is sometimes overlooked in the literature but will play a crucial role in our work. It depicts an additive contribution to the up Yukawa coupling obtained by closing all the lines in the 't Hooft vertex apart from those corresponding to $u$ and $u^c$.    Using the   arguments of App.~\ref{parametric}, and \eq{param} in particular, we   estimate  this contribution to be,
\bea
\Delta y^{\rm SI}_u \sim (16 \pi^2 \xi_I^{(3)}) \left(\dfrac{2\pi}{\alpha(M)}\right)^{6} e^{-\frac{2\pi}{\alpha(M)}}\left(\prod_i \frac{m_i}{M}\right)  y_d e^{-i (a/f_a+\bar{\theta})},
\label{upsi}
\eea
where $\alpha=g^2_s/(4 \pi)$ and $\xi_I^{(3)}=\xi^{(3)}_I (N_F,N_S)$ is a function of the number of colored Dirac fermions in the theory,  $N_F=n_F+2$, and the number of colored  complex scalars,  $N_S$, which is zero in this case. As discussed in App.~\ref{parametric}, this function arises from the integral over instanton sizes and is given by,
\begin{equation}
   \xi_I^{(3)} (N_F,N_S) = 2.5\times 10^{-3}\, e^{0.0625 \,\left(2N_F - N_S\right)}.
   \label{xi}
\end{equation}
Using \eq{xi}  we find that for $n_F=15$ the factor $16 \pi^2 \xi_I^{(3)}$ is ${\cal O}(1)$. One can similarly obtain the additive contribution to the down Yukawa,
\bea
\Delta \hat{y}_d^{\rm SI} \sim (16 \pi^2 \xi_I^{(3)}) \left(\dfrac{2\pi}{\alpha(M)}\right)^{6} e^{-\frac{2\pi}{\alpha(M)}}\left(\prod_i \frac{m_i}{M}\right)  \hat{y}_u e^{-i (a/f_a+\bar{\theta})}.
\label{upsi}
\eea
For $\hat{y}_u \ll {y}_d$ this contribution provides a small correction to the down Yukawa coupling but, as we will soon see, it cannot be neglected while computing the IR contribution to the axion potential.

As shown in    Fig.~\ref{toysib}, we obtain a contribution to the axion  potential   by closing all the lines of the  't Hooft vertex. Using \eq{param}  we obtain the following parametric estimate for this contribution, 
\bea
V_{\rm SI}(a)\sim  - \xi_{I} \left(\dfrac{2\pi}{\alpha(M)}\right)^{6} e^{-\frac{2\pi}{\alpha(M)}} \left(\prod_i \frac{m_i}{M}\right)  \frac{ \hat{y}_u y_d}{16 \pi^2} M^4 {\rm cos}\left(\frac{a}{f_a} + \bar{\theta}\right)
\label{potsi}
\eea
For both the estimates in \eq{potsi} and \eq{upsi} we have assumed that the dominant contribution comes from instantons of size $\rho\sim 1/M$, see App.~\ref{parametric}.

\subsection{The general lagrangian just above the QCD scale}

At scales below, $M$,  we can incorporate the small instanton contributions  of \eq{potsi} and \eq{upsi} into the following lagrangian, 
\bea
\mathcal{L} &\supset {}&
\frac{1}{2}\partial_\mu a\,\partial^\mu a
+\frac{g_s^2}{16\pi^2}
\left(\frac{a}{f_a}+\bar{\theta}\right)
\operatorname{Tr}G_{\mu\nu}\widetilde G^{\mu\nu}
+\hat{m}_{a,{\rm SI}}^2 f_a^2
\cos\left(\frac{a}{f_a}+\bar{\theta}\right)
\nonumber\\
&-&\Biggl[
\left(\hat y_u+\delta y_u^{\mathrm{SI}}
e^{-i(a/f_a+\bar{\theta})}\right)
\frac{h}{\sqrt{2}}\,u\,u^c
+
\left( y_d+\delta \hat{y}_d^{\mathrm{SI}}
e^{-i(a/f_a+\bar{\theta})}\right)
\frac{h}{\sqrt{2}}\,d\,d^c
+{{\rm h.c.}}
\Biggr].
\label{lint}
  \eea
where the only CP-violating phase  is $\bar{\theta}$ and,
\bea
\hat{m}_{a,{\rm SI}}^2&\sim&\xi_{I} \left(\dfrac{2\pi}{\alpha(M)}\right)^{6} e^{-\frac{2\pi}{\alpha(M)}}\left(\prod_i \frac{m_i}{M}\right)  \frac{\hat{y}_u}{4\pi }\frac{y_d}{4\pi }\frac{M^4}{f_a^2} ,\nn\\
\delta y_u^{\rm SI} &\sim & (16 \pi^2 \xi_{I}) \left(\dfrac{2\pi}{\alpha(M)}\right)^{6} e^{-\frac{2\pi}{\alpha(M)}} \left(\prod_i \frac{m_i}{M}\right) y_d\nn\\
\delta \hat{y}_d^{\rm SI} &\sim & (16 \pi^2 \xi_{I}) \left(\dfrac{2\pi}{\alpha(M)}\right)^{6} e^{-\frac{2\pi}{\alpha(M)}} \left(\prod_i \frac{m_i}{M}\right) \hat{y}_u.
\label{instacon}
\eea
The hats on $\delta \hat{y}_d^{\rm SI}$ and $\hat{m}_{a,{\rm SI}}^2$ emphasize that these contributions vanish in the symmetry-restoring limit $\hat{y}_u\to 0$.
Note that in \eq{lint} we have not explicitly included the mass terms   for the vector-like fermions,
\bea
{\cal L}_F = - \sum _i m_i F_i F^c_i+{\rm h.c.}
\eea
that have all been chosen to be real. As we move to scales below the mass of the lightest vectorlike fermion, $\mu \ll m_i^{\rm min}$, the $F_i$ are integrated out, leaving only the terms involving $u$, $u^c$, $d$ and $d^c$ in the Lagrangian of \eq{lint}. Between the scale $m_i^{\rm min}$ and the QCD scale  $\Lambda_{\rm QCD}$, the effective Lagrangian retains the form given in \eq{lint}. In particular, in the limit that, $\hat{y}_u,\; \delta \hat{y}_d^{\rm SI}$ and $\hat{m}_{a,{\rm SI}}^2$, all vanish we recover the symmetry in \eq{global1}. This ensures that these three parameters cannot be generated via renormalisation group (RG) flow from the two other couplings,   $\delta y_u^{\rm SI}$ and $y_d$. Notice that  these three 
are precisely the parameters that are proportional to, $\hat{y}_u$, and are small if  $\hat{y}_u$ is small; the above argument shows that this feature is preserved by RG flow.

While we have used the toy model in Sec.~\ref{toy} to derive the lagrangian in \eq{lint}, it is in fact more general.  As we will soon see,  the more realistic models of     Sec.~\ref{product} and~\ref{composite}   also have effects analogous to the  small instanton contributions in Fig.~\ref{toysi}. If we focus on only terms involving the first generation of fermions and scales just above the QCD scale, $\mu \gtrsim \Lambda_{\rm QCD}$, both these models    map to the lagrangian  in \eq{lint}  but with different  model dependent expressions for $\hat{m}_{a,{\rm SI}}^2,\; \hat{y}_u,\;\delta y_{u}^{\rm SI}$ and  $\delta \hat{y}_d^{\rm SI}$. Furthermore, even in these models, the small instanton contribution to the axion mass is proportional to a small parameter analogous to $\hat{y}_u$.

\subsection{ The general lagrangian at low scales, $\mu \lesssim \Lambda_{\rm QCD}$}
\label{chiral}
Depending on the value of $\hat{y}_u$ the small instanton contribution to the axion mass   in \eq{instacon} can be either greater than $\Lambda_{\rm QCD}$ or smaller. If  $\hat{m}_{a,{\rm SI}}^2\gg\Lambda^2_{\rm QCD}$,  we can   integrate out   the axion  by substituting in the lagrangian, $a=-f_a \bar{\theta}$. The  low-energy lagrangian  below $\Lambda_{\rm QCD}$ then reduces to that of standard QCD with a vanishing strong CP phase---a case that has been analysed in detail in the literature.

Here we will analyse the case, $\hat{m}_{a,{\rm SI}}^2\ll\Lambda^2_{\rm QCD}$, in the two-flavor chiral lagrangian. In App.~\ref{app:B} we perform the same analysis in the three-flavor chiral lagrangian.  Apart from the axion, $a$, the degrees of freedom at the scale  $\mu \lesssim \Lambda_{\rm QCD}$ are $\pi^0,\; \pi^\pm$ and $\eta'$. These are embedded in the matrix, 
\begin{equation}
    U = e^{i\pi^a \sigma^a/f_{\pi} + i\eta'I/f_{\eta'}},
\end{equation}
 $\sigma_a$ being  the Pauli matrices. Note that the physical range of $\eta'/f_{\eta'}$ is $\left[0, 2\pi /F\right]$, $F$ being the number of light flavors (see for instance Ref.~\cite{Csaki_Gupta_Roy}).

The lagrangian in \eq{lint} matches to the following effective lagrangian below the QCD scale, 
\begin{align}
    \mathcal{L}_{\chi} &= \frac{f_{\pi}^2}{4}\mbox{ Tr }\partial_{\mu}U\partial^{\mu}U^{\dagger} + B_0f_{\pi}^3 \mbox{ Tr}(M_QU + M_Q^{\dagger}U^{\dagger}) - F(e^{-i\left(\frac{ a}{f_a} +\bar{\theta}\right)} \mbox{ det}\,U)  \nonumber\\
    &+  \frac{1}{2}\partial _\mu a \partial^\mu a+ \hat{m}_{a,{\rm SI}}^2 f_a^2 \text{ cos}\left(\frac{a}{f_a} + \bar{\theta}\right). \label{lchi}
\end{align}
Here, 
\begin{equation}
    M_Q = \text{diag}\left\{
        \hat{m}_u + m_u^{\rm SI} \,e^{-i(a/f_a + \bar{\theta})}\,,\;m_d+ \hat{m}_d^{\rm SI} \,e^{-i(a/f_a + \bar{\theta})}\right\},
\end{equation}
with  $\hat{m}_u = \hat{y}_u v/\sqrt{2}$, $m_u^{\rm SI}=\delta y_u^{\rm SI} v/\sqrt{2}$, $\hat{m}_d^{\rm SI}=\delta \hat{y}_d^{\rm SI} v/\sqrt{2}$,  $m_d =y_d v/\sqrt{2}$;  $v=246$ GeV is the Higgs VEV. In addition to the above lagrangian terms the Weiss-Zumino-Witten (WZW)~\cite{Wess:1971yu,Witten:1983tw} term  generates a pion photon coupling given by,
\bea
\mathcal{L}_{WZW} = -\frac{3 \alpha_{em}}{4\pi} \text{ Tr}\left[Q^2 \left(\frac{\eta'}{f_{\eta'}}\mathbb{I}_2 + \frac{\pi^0}{f_{\pi}}\sigma^3\right)\right] F^{\mu\nu}\tilde{F}_{\mu\nu}.
\eea
where $Q$ is the quark charge matrix given by $Q = \text{diag}(2/3,\,-1/3)$. 

The function, $F(e^{-i\left(a/f_a +\bar{\theta}\right)}\det U)$, encodes  the information about the low energy breaking of  axial symmetry due to the anomaly and provides the dominant contribution to the $\eta'$ mass, see for instance Ref.~\cite{Csaki_Gupta_Roy}. The form of the argument of this function can be understood from the fact that the lagrangian must respect the global symmetry in \eq{global1} for $\hat{y}_u\to 0$ and hence only the invariant combination, $e^{-i(a/f_a + \bar{\theta})}\det U$, can appear in this term. The simplest possibility,
 \bea
 F(e^{-i(a/f_a + \bar{\theta})}\mbox{ det}\,U)=-A f_\pi^4 e^{-i\left(\frac{ a}{f_a}+\bar{\theta}\right)} \det U  +{\rm h.c.}
 \eea
 is expected to arise from instanton effects at the QCD scale. Using large-$N$ arguments it was argued in Ref.~\cite{ven1,ven2, witten}, however,  that  these instanton effects should be suppressed. Instead, Ref.~\cite{witten} proposed the following   branched potential for the $\eta'$, 
 \bea
 F(e^{-i(a/f_a + \bar{\theta})}\mbox{ det}\,U)=\frac{a f_\pi^2}{ 2N_c} \left[ i\ln \left(e^{-i(a/f_a + \bar{\theta})}\det U \right)\right]^2+\cdots
 \eea
where $N_c=3$ is the number of colors.\footnote{In Ref.~\cite{Csaki_Gupta_Roy}, motivated by calculations in a supersymmetric extension of QCD,  the following form for this function was proposed, $F(e^{-i(a/f_a + \bar{\theta})}\mbox{ det}\,U)=-{N_c \Lambda^2_{\rm QCD}}{f_\pi^2}\left(e^{-i(a/f_a + \bar{\theta})}\det U \right)^{1/N_c}+\rm{h.c.}$  This form also satisfies the large-$N$ arguments of Ref.~\cite{ven1, ven2, witten}.} For our purposes all these possibilities result in the same lagrangian once, $\eta'$ is integrated out.  This is because all these possibilities result in an $\eta'$ potential that is of the same form near the minima,
\bea
V(\eta')\sim m^4 \left(\frac{2 \eta'}{f_\eta'}-\frac{a}{f_a}-\bar{\theta}\right)^2,
\eea
$m$ being a dimensionful constant. Integrating out $\eta'$,
\begin{equation}\label{inteta}
   \eta'  = \frac{1}{2} f_{\eta'}\,\left(\frac{a}{f_a}+\bar{\theta}\right),
\end{equation}
and  substituting it in \eq{lchi}  gives us a potential for the pion and axion that is independent of the specific form of $F(e^{-i(a/f_a + \bar{\theta})}\mbox{ det}\,U)$. We obtain,
\bea\label{eq:QCD_potential}
    V(a, \pi^0) &=& -m_{\pi}^2f_{\pi}^2 \sqrt{1-\frac{4(\hat{m}_u + \hat{m}_d^{\rm SI})(m_d + m_u^{\rm SI})}{(\hat{m}_u + m_d + m_u^{\rm SI} + \hat{m}_d^{\rm SI})^2}\text{ sin}^2\left(\frac{a}{2f_a} + \frac{\bar{\theta}}{2}\right)} \text{cos}\left(\frac{\pi^0}{f_{\pi}} - \phi_a\right)\nn\\
    &-& \hat{m}_{a,{\rm SI}}^2 f_a^2 \cos\left(\frac{a}{f_a}+\bar{\theta}\right)
\eea
where, 
\begin{equation}
    \text{tan}\,\phi_a \equiv -\frac{(\hat{m}_u + \hat{m}_d^{\rm SI})- (m_d + m_u^{\rm SI})}{\hat{m}_u + m_d + m_u^{\rm SI} + \hat{m}_d^{\rm SI}} \text{ tan}\,\left(\frac{a}{2f_a} + \frac{\bar{\theta}}{2}\right)\;.
\end{equation}
Finally integrating out $\pi^0$,
\bea
\frac{\pi^0}{f_{\pi}}= \phi_a,
\label{intpion}
\eea 
gives the effective axion potential, 
\begin{equation}\label{fullpot}
    V(a) =  -m_{\pi}^2f_{\pi}^2 \sqrt{1-\frac{4(\hat{m}_u + \hat{m}_d^{\rm SI})(m_d +m_u^{\rm SI})}{(\hat{m}_u + m_d + m_u^{\rm SI} + \hat{m}_d^{\rm SI})^2}\text{ sin}^2\left(\frac{a}{2f_a} + \frac{\bar{\theta}}{2}\right)}-\hat{m}_{a,{\rm SI}}^2 f_a^2 \cos\left(\frac{a}{f_a}+\bar{\theta}\right).
\end{equation}
This potential gets minimised at 
\bea
\frac{\langle a\rangle}{f_a} + \bar{\theta}=0
\eea
thus solving the strong $CP$ problem.

Expanding around this minimum to quadratic order   gives the axion mass, 
\begin{equation}
m_a^2  = \frac{(\hat{m}_u + \hat{m}_d^{\rm SI})\,(m_u^{\rm SI} +m_d)}{(\hat{m}_u +m_u^{\rm SI} +m_d + \hat{m}_d^{\rm SI})^2}\frac{m_{\pi}^2 f_{\pi}^2}{f_a^2}  +   \hat{m}_{a,{\rm SI}}^2  \;
\label{axionmass}
\end{equation}
where the first term comes from the QCD potential~(\ref{eq:QCD_potential}) in IR while the second term arises from small-instanton effects in UV. Notice that for the toy model of Sec.~\ref{toy}, $\hat{m}_u, \hat{m}_d^{\rm SI}$ and $\hat{m}_{a,{\rm SI}}^2$, all vanish as $\hat{y}_u \to 0$. This implies that the axion mass can be made arbitrarily small by choosing a small enough, $\hat{y}_u$. The fact that each of the terms in \eq{axionmass} can be arbitrarily small is a feature of all the models proposed in this work.

In App.~\ref{app:B} we compute the axion mass in the chiral lagrangian with three light flavors. We find that the expression for the IR contribution vanishes as $\hat{y}_u\to 0$ even for the 3-flavor case and the 3-flavor expression reduces to the one in \eq{axionmass} if we ignore  ${\cal O}(m_{d}/m_s)$ effects.  

\paragraph{Axion-photon coupling} If the PQ  symmetry has an electromagnetic anomaly the axion also has a coupling to photons above the QCD scale,
\bea
  \mathcal{L}_{a\gamma\gamma} (\mu \gtrsim \Lambda_{\rm QCD}) =\frac{\alpha_{em}}{8\pi}\frac{E}{N}\frac{a}{f_a} F^{\mu\nu} \tilde{F}_{\mu\nu}
  \eea
where $E$ and $N$ are the electromagnetic and color anomaly coefficients of the PQ symmetry. This coupling is modified when we integrate out $\eta'$ and the pion using \eq{inteta} and \eq{intpion},
\begin{equation} \label{eq:aFF}
    \mathcal{L}_{a\gamma\gamma} (\mu \lesssim \Lambda_{\rm QCD}) = \frac{\alpha_{em}}{8\pi f_a} \left(\frac{E}{N} - \frac{2}{3}\frac{4m_d + \hat{m}_u +4m^{\text{SI}}_u + \hat{m}_d^{\rm SI}}{m_d + \hat{m}_u + m^{\text{SI}}_u + \hat{m}_d^{\rm SI}}
    \right)a F^{\mu\nu} \tilde{F}_{\mu\nu}.
\end{equation}
Note that while the second term within the parenthesis   is different from the usual one~\cite{QCD_axion_precisely} it is still an ${\cal O}(1)$ number.

\subsection{Necessary conditions for obtaining a light axion}
\label{conditions}
Both the contributions to the mass of the axion in  \eq{axionmass} are proportional to $\hat{y}_u$. This is ultimately related to the fact that in the limit, $\hat{y}_u \to 0$,  we recover the symmetry in \eq{global1} and the axion truly becomes a massless goldstone boson. 

The existence of an approximate symmetry that protects the axion mass is a common feature of all the models we are going to consider. Hence, in all these models there is a small parameter, $\hat{y}$, such that as $\hat{y} \to 0$ all the contributions to the axion mass vanish, that is:  (1) $\hat{m}_{a,{\rm SI}}^2\to 0$, (2) $\delta \hat{y}_d^{\rm SI}\to 0$, and (3) $\hat{y}_u(\mu)\to 0$ at  scales just above $\Lambda_{\rm QCD}$. In the toy model of Sec.~\ref{toy}, $\hat{y} = \hat{y}_u (M)$.   Note that in this model   a small value of $\hat{y}_u$ at the scale  $M$ implies a small value at lower scales.

The IR contribution to the axion mass, i.e. the first term in \eq{axionmass}, is proportional to $\hat{m}_u=\hat{y}_u v/\sqrt{2}$. In order for this contribution to be suppressed we require a  small, $\hat{y}_u \ll1$. As the up quark Yukawa just above QCD scale is given by the sum, $\hat{y}_u   + \delta y_u^{\rm SI}$,\footnote{Here the phase of the small instanton contribution in \eq{lint} does not appear because the axion relaxes to a minimum where , $\bar{\theta}+  {\langle a \rangle}/{f_a}=0$} this in turn requires that  the small instanton contribution should almost entirely account for the observed up Yukawa coupling. Lattice simulations  require the following value for the ratio of the up and down Yukawas  in the $\overline{\rm MS}$ scheme\footnote{Note that the Kaplan-Manohar ambiguity~\cite{km} that affects the chiral lagrangian (see also Ref.~\cite{bns, troy}) does not apply to  lattice simulations as they   directly connect observables to these high scale   Yukawa couplings.},
\bea\label{conditionratio}
\frac{y_u^{\overline{\rm MS}}(2{~\rm GeV})}{y_d^{\overline{\rm MS}}(2{~\rm GeV})}= 0.46(2).
\eea
In the limit the small parameter, $\hat{y}\to 0$,  we have, $\hat{y}_u \to 0$ and  $\delta \hat{y}_d^{\rm SI}\to 0$. Thus, in this limit we have for the up and down Yukawa couplings in \eq{lint}, 
\bea
\hat{y}_u   + \delta y_u^{\rm SI} \to \delta y_u^{\rm SI}\nn\\
{y}_d   + \delta \hat{y}_d^{\rm SI} \to y_d.
\eea
The condition in \eq{conditionratio} therefore gives, 
\bea
\frac{\delta y_u^{\rm SI} (2{~\rm GeV})}{y_d(2{~\rm GeV})} = 0.46,
\label{value}
\eea
 to a very good approximation for $\hat{y} \ll1$. 

 In the toy model of Sec.~\ref{toy}, we do not require the Yukawa couplings to match the observed  values. This allows us to take $\hat{y}_u$ arbitrarily small and thus have the QCD axion   arbitrarily light.

The condition  in \eq{value} is hard to realise in most UV models. This is because in the UV the `t Hooft vertex will contain  all the SM fermions and one expects the following suppression factor in  $\delta y_u^{\rm SI}$,
\bea
\varepsilon_y=\frac{y_d}{4 \pi}\frac{y_c}{4 \pi}\frac{y_s}{4 \pi} \frac{y_b}{4 \pi} \frac{y_t}{4 \pi}\sim 10^{-19},
\label{epsy}
\eea
where we have used the value of the Yukawa couplings at  the scale $\mu \sim 100$ TeV from Ref.~\cite{antusch2026}. Therefore in order to have a light QCD axion,  the UV theory is required to not only  overcome the suppression of small instanton effects due to a small $\alpha_s$ but  also   the suppression due to the above factor $\varepsilon_y$. It is therefore necessary to consider models  that modify not only QCD but also the SM flavor sector in the UV such as those  in Sec.~\ref{product} and Sec.~\ref{composite}.  In particular, the suppression  factor $\varepsilon_y$ may be absent in  models that solve the SM flavor puzzle  as in such scenarios  the  low-energy smallness of the   Yukawa couplings  of SM quarks    may arise from a UV theory with ${\mathcal O}(1)$ parameters. The model in Sec.~\ref{composite}, in fact, realises  exactly such a scenario as the SM flavor puzzle is addressed by utilising partial compositeness.

\section{Light QCD axion in models with modified gauge groups}
\label{product}
We now discuss a model where the QCD gauge group emerges in the IR as a subgroup of a larger product group, $SU(3)_1 \times SU(3)_2 \times SU(3)_3 \times \cdots SU(3)_ N$. The gauge coupling of the individual, $SU(3)_i$, groups can be much larger than the QCD gauge coupling which makes the small instantons of these groups unsuppressed in the UV. As far as the suppression factor $\varepsilon_y$ in \eq{epsy} is concerned, it is absent because not all the SM quarks are charged under the same $SU(3)_i$ group.

\subsection{Model}
\begin{figure}[t]
    \centering
    \includegraphics[width=0.8\linewidth]{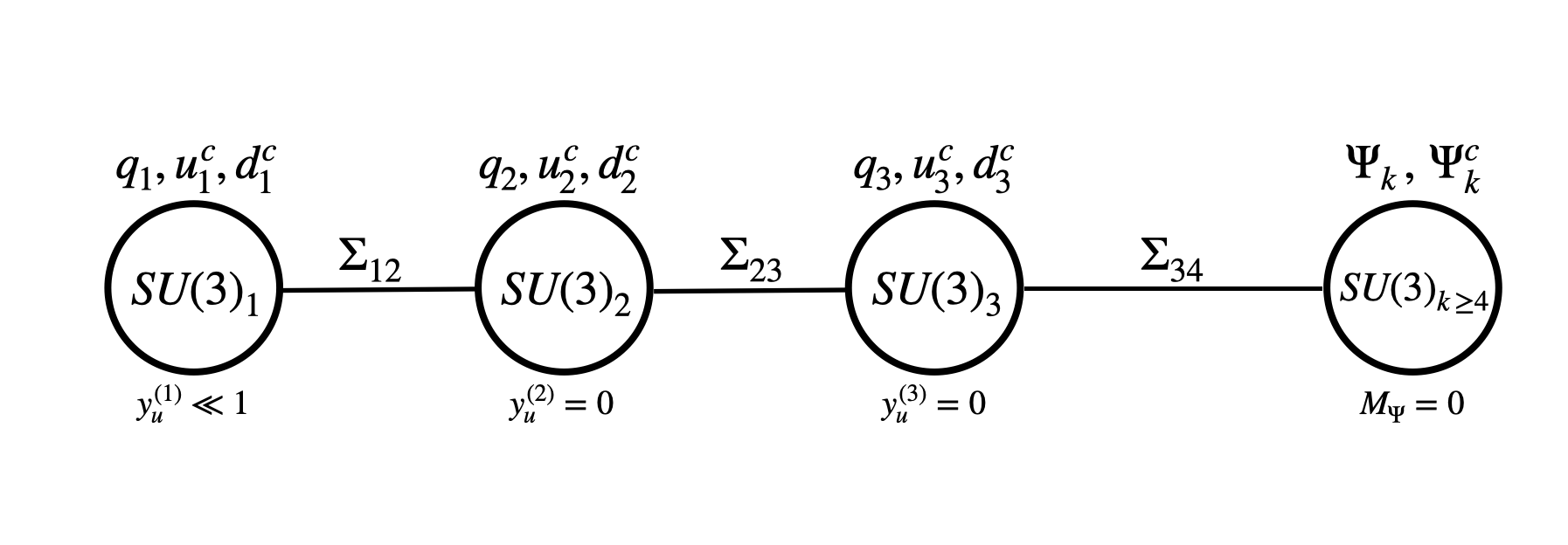}
    \caption{Schematic representation of the model of Sec.~\ref{product} in theory space.}
    \label{schematic}
\end{figure}
We consider a model similar to that proposed in Ref.~\cite{Agrawal:massless_up}.  We present a schematic representation of the model in Fig.~\ref{schematic} and summarise the full particle content and charge assignments  in Table~\ref{tab:UV_particles}. The lagrangian of our model is invariant under a product of $SU(3)$ gauge groups, $SU(3)_1 \times SU(3)_2 \times SU(3)_3 \times \cdots SU(3)_N$, and contains the following terms,
\begin{align}
    {\mathcal{L}} &\supset -\frac{1}{2}\sum_{k=1}^N \text{Tr}\, (G_k^{\mu\nu}G_{k,\,\mu\nu})   - V(\Sigma).  \label{eq:UV_L}
\end{align}
Here $V(\Sigma)$ is the potential involving the scalar link fields, $\Sigma_{i\,i+1}$, that transform in the bifundamental $(3,\,\bar{3})$ representation under the adjacent $SU(3)_i \times SU(3)_{i+1}$ gauge groups.  Each of the scalar link fields, $\Sigma_{i\,i+1}$, develops a vacuum expectation value (VEV), $v_\Sigma\sim M/g_\Sigma$, from a standard Higgs-like renormalizable potential where $g_\Sigma$ is a coupling factor. This spontaneously breaks the product gauge group down to the diagonal subgroup $SU(3)_D$,
\bea
SU(3)_1 \times SU(3)_2 \times SU(3)_3 \times \cdots SU(3)_N \longrightarrow SU(3)_D\equiv SU(3)_{\rm QCD}.
\eea
 The surviving diagonal subgroup, is identified with the  QCD group of the SM. The coupling of  $SU(3)_{\rm QCD}$  at the scale $M$ is given at tree-level by the relation,
\begin{equation} 
    \frac{1}{\alpha_s} = \frac{1}{\alpha_{s_1}} + \frac{1}{\alpha_{s_2}} + \frac{1}{\alpha_{s_3}}+\sum_{i=4}^N\frac{1}{\alpha_{s_i}}.
\end{equation}
This matching relation implies that $\alpha_s(M)$ can remain small even if the individual gauge couplings, $\alpha_{s_i}$, of the product group are larger. As the non-perturbative instanton effects scale as $\sim e^{-2\pi/\alpha_{{s_i}}}$, they can then have a sizeable contribution just above the scale $M$.

The fermionic matter content of the model is as follows: the first, second and third generations of SM quarks are charged under the first, second and third  $SU(3)$ groups, respectively.  We also assume the existence of  massless vector-like fermions charged under each of the other $SU(3)$ gauge groups  so that the $\theta$-angles for each of these groups can also be rotated away. The   lagrangian terms   relevant for the axion mass and the strong CP problem are as follows,
\bea \label{productflv}
{\cal L}\supset \frac{g_1^2}{16\pi^2}\left(\frac{a}{f_a} + \bar{\theta} \right)\text{Tr}\,G_1^{\mu\nu}\tilde{G}_{1,\,\mu\nu} -
 \hat{Y}_u\, q_1Hu_1^c - Y_d\, q_1H^{\dagger} d_1^c    - Y_c\, q_2H u_2^c - Y_t\, q_3H u_3^c, \nonumber\\
 \eea   
  where the quark fields above are in the flavor basis.  While writing the above lagrangian terms we have performed chiral rotations of the up and down  quarks of the $SU(3)_1$ sector to rotate the phases of the corresponding Yukawa couplings into the first term; the Yukawa couplings, $Y_{u,d}$, are thus real and positive. As far as the second and third generation fermions are concerned,  we   consider the technically natural scenario where the strange and bottom quarks have no Yukawa couplings. This allows us to choose, without any loss of generality,  a real $Y_{c,t}$ and to also rotate away the $\theta$-angles for the $SU(3)_2$ and $SU(3)_3$ groups. We show in App.~\ref{app:flavorCP} that small enough  values for strange and bottom Yukawa couplings  do  not spoil our mechanism. 

 In this model the small parameter discussed in Sec.~\ref{conditions}, $\hat{y} =\hat{Y}_u$. In the limit $\hat{Y}_u\to 0$ we recover the non-anomalous symmetry,
  \bea
  u^c \to e^{i\alpha} u^c ,\,~~~~ \frac{a}{f_a}\to \frac{a}{f_a}+\alpha.
\label{global2}
\eea
This symmetry protects the axion mass and  in the limit, $\hat{Y}_u\to 0$,  all contributions to the axion mass vanish.

\begin{table}[t]
\renewcommand{\arraystretch}{1.2}
    \centering
    \begin{tabular}{c|cc}
    \hline
        Particles & $SU(3)_i$ & $SU(3)_j$\\
    \hline
    \hline
    $G_i$ & 8 & 1 \\
    \hline
     $q_i$ & $3$ & 1 \\
     $u_i^c$ & $\bar{3}$ & 1 \\
     $d_i^c$ & $\bar{3}$ & 1 \\
     \hline
     $\Sigma_{ij}$ & $3$ & $\bar{3}$ \\
     \hline 
     $H$ & 1 & 1 \\
     \hline
    \end{tabular}
    \caption{Particle content and  charge assignments in the model of Sec.~\ref{product}. Here, $q_i$ refers to the left-handed SM quarks of $i^{th}$ generation and $u_i^c\,(d_i^c)$ are the right-handed up-(down-) type SM quarks of $i^{th}$ generation.}
    \label{tab:UV_particles}
\end{table}

So far we have not discussed how the different quark generations can mix and reproduce the  pattern of SM quark masses and mixings. In Ref.~\cite{Agrawal:massless_up} this is realised by higher dimensional mixing terms between the different generations involving the $\Sigma_{ij}$ fields; we adopt the same mechanism and refer to~\cite{Agrawal:massless_up} for  a detailed discussion. It has also been shown in Ref.~\cite{Agrawal:massless_up} that flavor bounds require a lower bound,  $M \gtrsim 1000$ TeV. As far as terms involving the first generation of quarks are  concerned, this theory maps to the lagrangian in \eq{lint}  for scales $\Lambda_{\rm QCD}\lesssim \mu \lesssim M$, provided we keep the Yukawa couplings of all the other quarks to be real in the mass basis.  As   \eq{lint} is written in the mass basis we must carry out the appropriate flavor rotations in order to match \eq{productflv} to \eq{lint}.  Following Ref.~\cite{Agrawal:massless_up} we have, 
\bea
\hat{y}_u=\hat{Y}_u c_{\theta_c},~~~~~~~~~~~~~~~{y}_d={Y}_d. 
\label{flvmix}
\eea
Here $c_{\theta_c}=\cos \theta_c$, $\theta_c$ being the Cabbibo angle. We now proceed to  the computation of the small instanton contributions to $\hat{m}_{a,{\rm SI}}^2$ and $\delta y_u^{\rm SI}$ in this model.

\subsection{Small Instanton Contributions}
\begin{figure}[t]
\begin{subfigure}{0.5\textwidth}
\hspace{0.5cm}
\centering
\includegraphics[width=0.7\linewidth]{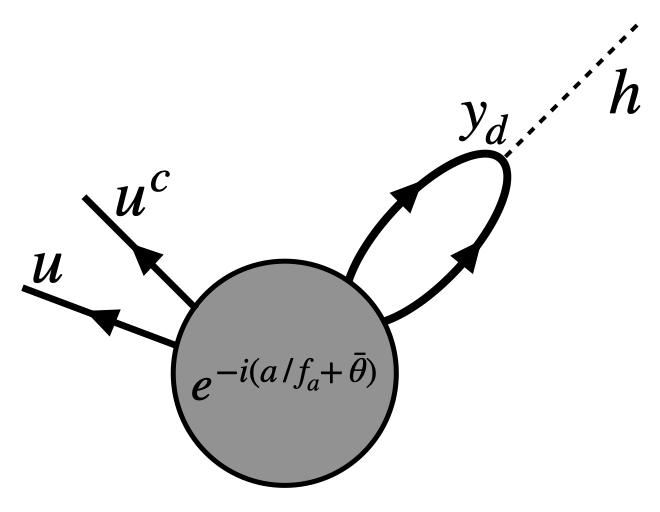} 
\caption{}
\label{fig:up_quark_model1}
\end{subfigure}
\begin{subfigure}{0.5\textwidth}
\hspace{-0.4cm}
\centering
\includegraphics[width=0.65\linewidth]{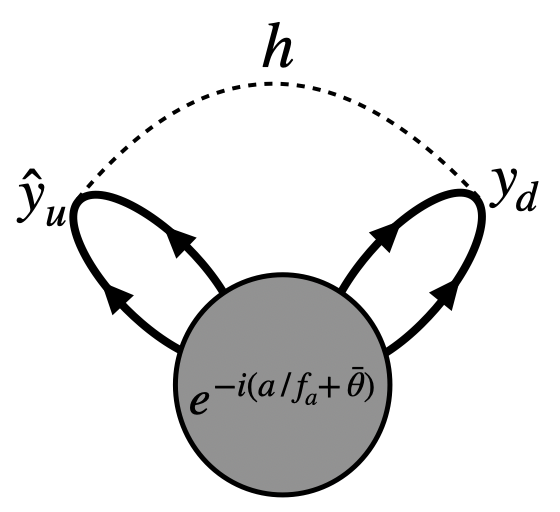}
\caption{}
\label{fig:axion_potential_model1}
\end{subfigure}
\caption{Small instanton contributions to (a) the up-quark Yukawa, and (b)  the axion potential in the model of Sec.~\ref{product}. A small instanton correction to the down quark Yukawa is obtained from a diagram with the up and down quark lines interchanged  and the coupling $y_d$ interchanged with $\hat{y}_u$.}
\label{fig:model1}
\end{figure}

At scales, $\mu \gtrsim M$, the  instantons  of  the $SU(3)_i$ groups become unsuppressed. These contributions are cut-off at the Higgsing scale $M$ and are thus dominated by small instantons of sizes $\rho \lesssim M^{-1}$.
The two kinds of small instanton contributions discussed in Sec.~\ref{toy} must both be considered to correctly compute the axion mass. 

\paragraph{Contribution to   up-type Yukawa couplings}  The small instanton contribution to up-quark mass is obtained by closing all the lines except the up-quark lines as   shown in Fig.~\ref{fig:up_quark_model1}.  The computation of this effect was  performed using  the dilute instanton gas approximation  and some other approximations in Ref.~\cite{Agrawal:massless_up}. We review this computation in App.~\ref{parametric} and obtain (see \eq{upproduct}),
   \bea\label{eq:up_Yukawa}
\frac{\delta y_u^{\rm SI}}{y_d} = \frac{4\pi^2 c_{\theta_c}}{3b}\xi_I^{(3)}(2,  3)\,\Gamma\left(7, \,2\pi/\alpha_{s_1}(M)\right)\, 
\eea
 where,
 \bea
\xi_I^{(3)} (2, 3)= 2.7 \times 10^{-3},
\eea
and we have replaced $Y_d $ by $y_d$ using  \eq{flvmix}.  Here  $b=55/6$ is the coefficient of the beta function of $SU(3)_1$. The small instanton contribution in \eq{upproduct} generates a correction to the `11' element of the up quark Yukawa matrix, $\delta Y_u^{\rm SI}$,  in the flavor basis. Going from the flavor basis to the  up Yukawa in the mass basis introduces the factor of $c_{\theta_c}$ above, see \eq{flvmix} and Ref.~\cite{Agrawal:massless_up}.  A similar diagram gives the small instanton contribution to the `22' element of the down Yukawa matrix, see \eq{appdownsi}; after appropriately including flavor rotation factors from \eq{flvmix} we get,  
\bea\label{eq:dwn_Yukawa}
\delta \hat{y}_d^{\rm SI} = \frac{4\pi^2 }{3b \,c_{\theta_C}}\xi_I^{(3)}(2,  3)\,\Gamma\left(7, \,2\pi/\alpha_{s_1}(M)\right) \hat{y}_u\,. 
\eea

As discussed in Sec.~\ref{conditions}, a light QCD axion is possible provided we can take $\hat{y}_u \ll {\delta y_u^{\rm SI}}$ which in turn  requires that $ {\delta y_u^{\rm SI}}/{y_d}=0.46$ at low scales, see \eq{conditionratio}. While \eq{eq:up_Yukawa} provides $ {\delta y_u^{\rm SI}}/{y_d}$ at  the high-scale, $\mu \sim M$, this ratio is stable under RG to a very good approximation\footnote{This follows from the fact that the ratio of up and down Yukawas is approximately stable under RG in the SM because the dominant  contributions to RG  flow proportional to $g_s^2$ and $g_2^2$ are identical for up and down quarks where $g_2$ is the $SU(2)_L$ gauge coupling. The main difference is due to the hypercharge contribution which is very small so that the ratio $y_u/y_d$ in the SM equals 0.46 even at $M=1000$ TeV. The RG running of $\delta y_u^{\rm SI}$ is expected to be the same as that of the up Yukawa in the SM as we can treat the axion in  the corresponding lagrangian term in \eq{lint} as a background field.} so that we can impose the requirement in \eq{conditionratio} directly on the expression in \eq{eq:up_Yukawa}. This gives, 
\bea
\alpha_{s_1}(M) = 0.7.
\label{gauge1}
\eea

Small instanton effects at the second and third site in Fig.~\ref{schematic} also generate the strange and bottom Yukawa couplings. As discussed in Ref.~\cite{Agrawal:massless_up}, by appropriately choosing $\alpha_{s_2}(M)$ and $\alpha_{s_3}(M)$ we can similarly ensure that these Yukawa couplings  match their observed values. 

\paragraph{Contribution to axion potential} As shown in  Fig.~\ref{fig:axion_potential_model1},  the contribution of small instanton configurations to the axion potential can be obtained by completely closing all the fermionic legs of the `t Hooft vertex for the  $SU(3)_1$  group. In App.~\ref{parametric},  following Ref.~\cite{Csaki:many_heavy_axion}, we use the dilute instanton gas approximation and some other assumptions to obtain,
 \bea\label{eq:axion_potential}
 \hspace{-0.5cm}
    V_{\rm SI} (a)
    &=& -\left(\frac{2}{b-4}\right)\xi_I^{(3)} (2,3)  \left(\dfrac{2\pi}{\alpha_{s_1}(M)}\right)^{6} e^{-\frac{2\pi}{\alpha_{s_1}(M)}} \,\frac{\hat{y}_u y_d}{24\pi^2 c_{\theta_C}}M^4 \text{ cos}(a/f_a + \bar{\theta}),
    \eea
    where we have substituted the values of $\hat{Y}_u$ and $Y_d$ in \eq{massproduct} using \eq{flvmix}. This gives for the small instanton contribution to the axion mass,
\bea
\label{msiprod}
\hat{m}_{a,{\rm SI}}^2 &=& \xi_I^{(3)}(2,\, 3)  \left(\dfrac{2\pi}{\alpha_{s_1}(M)}\right)^{6} e^{-\frac{2\pi}{\alpha_{s_1}(M)}} \left(\frac{2}{b-4}\right)\frac{\hat{y}_u y_d}{24\pi^2 c_{\theta_C}f^2_a} M^4,
\eea
To obtain the full axion mass in the limit $\hat{y}_u \ll {\delta y_u^{\rm SI}}$  we  use \eq{axionmass}, \eq{gauge1} and \eq{msiprod},
\bea\label{axionmassproduct} 
m^2_a&\sim& (2 \times 10^{11} \sqrt{\hat{y}_u} \text{ meV})^2 \left(\frac{10^9 \text{ GeV}}{f_a}\right)^2 \left(\frac{M}{2000 \text{ TeV}}\right)^4,
\eea
where the contribution from the IR term  in \eq{axionmass}  is orders of magnitude smaller than that from \eq{msiprod}.

\begin{figure}[t]
    \centering
    \hspace{-1cm}
    \includegraphics[width=0.8\linewidth]{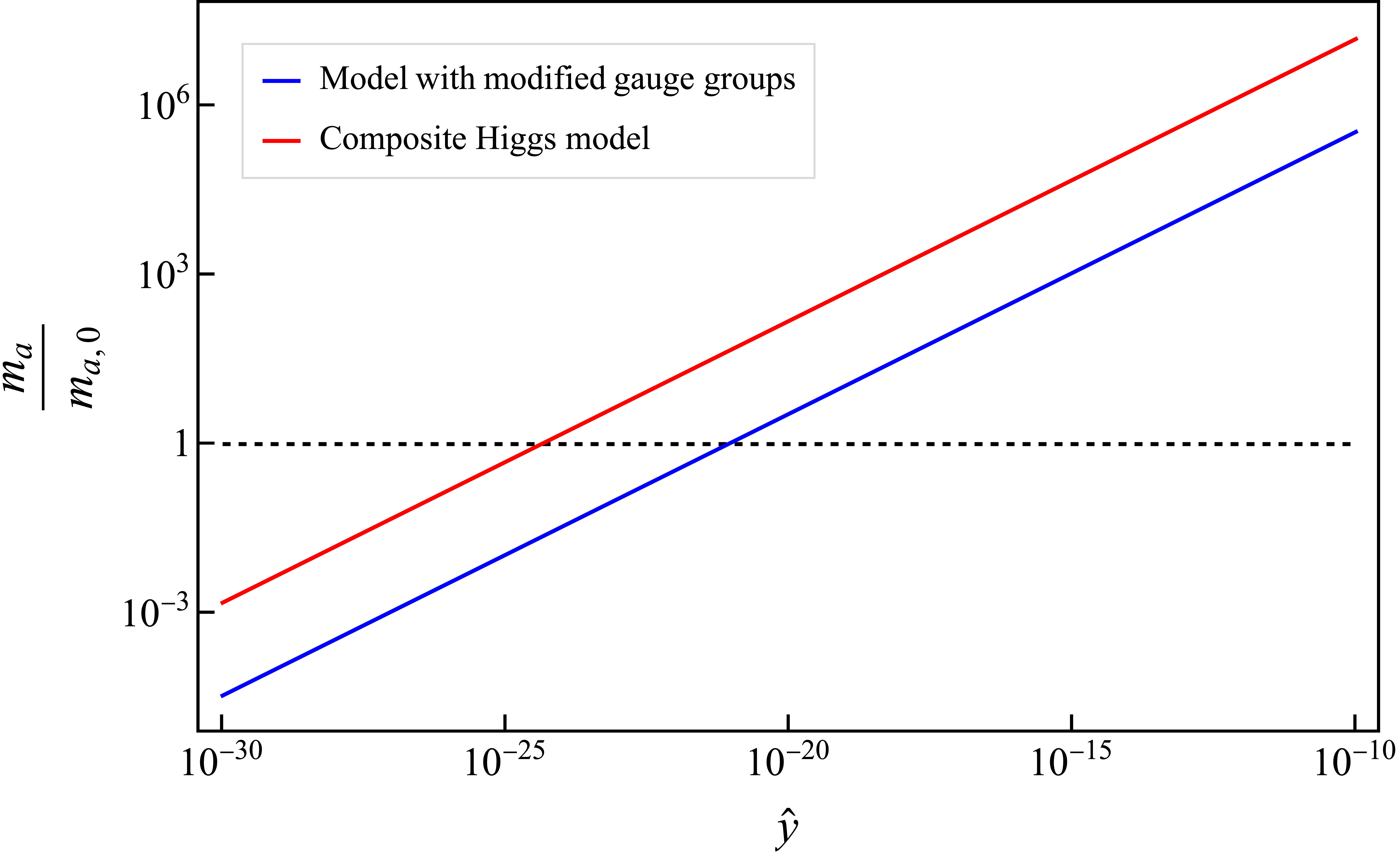}
    \caption{The ratio of the QCD axion mass to the standard expectation, $m_{a,0}$,  plotted as a function of the small parameter, $\hat{y}$, in the models of Sec.~\ref{product} and Sec.~\ref{composite}; the expression for the axion mass in these models is given, respectively, by \eq{axionmassproduct} and \eq{axionmasscomposite} where we have taken $M=2000$ TeV. The small parameter is given by $\hat{y}= \hat{Y}_u \approx \hat{y}_u$ ($\hat{y}=\hat{y}_{u_L}$) in the model of Sec.~\ref{product} (Sec.~\ref{composite}).}
    \label{fig:massplot}
\end{figure}

In Fig.~\ref{fig:massplot} we show how the axion mass varies with $\hat{Y}_u$ in this model. We see that it can be smaller than the standard expectation for $\hat{Y}_u\approx \hat{y}_u\lesssim 10^{-22}$ and is heavier otherwise. To obtain a very light axion we therefore require, $\hat{Y}_u$,  to be very small. While such small couplings are technically natural because setting them to zero restores a global symmetry as in \eq{global1}, exact global symmetries are not expected to survive quantum gravity~\cite{Harlow:2018jwu}. The model proposed here might, however, arise from a   larger framework  where the global symmetry accidentally survives up to a certain operator dimension once gauge symmetries--- possibly including    discrete ones---are imposed on the lagrangian. The breaking of the symmetry by higher dimensional Planck-suppressed operators  might then explain the smallness of $\hat{Y}_u$. 

Note that the above estimates are the leading order contributions to the lagrangian in \eq{lint}. There can be additional contributions due  to   effects involving the $\Sigma_{ij}$ fields that generate the observed CKM phase and mixings. In Ref.~\cite{Agrawal:massless_up} the quark $u^c$ does not have any couplings to $\Sigma_{ij}$ so that the symmetry in \eq{productflv}  remains intact even after the inclusion of these effects. In particular, the axion remains massless in the limit, $\hat{Y}_u \to 0$.   Higher order effects---including both  radiative effects and possible contributions due to a CP violating sector at higher scales---can, however, induce  new misaligned contributions to the axion potential that give a residual strong CP phase. As we discuss in App.~\ref{app:flavorCP} in the limit, $\hat{Y}_u \to 0$, the residual strong CP phase equals the value estimated in Ref.~\cite{Agrawal:massless_up} which is within experimental limits. We show in App.~\ref{app:flavorCP} that  the strong CP phase is expected to remain within experimental bounds   even after  $\hat{Y}_u$ is turned on.

\section{Light QCD axion in composite Higgs models}
\label{composite}

In this section we discuss how small instanton effects can become unsuppressed in composite Higgs models that realise partial compositeness.  The number of constituent colored  quarks forming the composite states  can be very high which results in QCD becoming strongly coupled in the UV. Moreover, the flavor suppression in \eq{epsy} can be overcome because these models can have an anarchic flavor structure in the UV with flavor hierarchies arising  only in the IR. The mass of a QCD axion  in such models  can therefore deviate from  standard expectations.

\subsection{Model}
 In composite Higgs models, the approximate global symmetry group $G_F$ of  a strongly interacting sector is spontaneously broken to a subgroup $H_F$ upon confinement.  Some of the associated pseudo Nambu-Goldstone bosons residing in the coset space, $G_F/H_F$, are identified as the Higgs doublet.  We will  consider the  model presented in Sec.V of  Ref.~\cite{RSG_Comp_Higgs}. This model is itself an extension of Ref.~\cite{Ferretti_2014} and considers   composite Higgs model where an  approximate  global $SU(5) \times U(1)_X$ symmetry   is spontaneously broken to $SO(5) \times U(1)_X$. This model also  contains a QCD axion and enhanced small instanton effects. Unlike Ref.~\cite{RSG_Comp_Higgs} that considers only the possibility of a heavy QCD axion, we show that the axion in this model can be  both lighter and heavier than  standard expectations.

 \begin{table}[t]
\begin{adjustbox}{max width=\textwidth}
    \centering
    \begin{tabular}{c|c|cccccccc}
         \hline
         \hline
         & $G_{HC}$ & \multicolumn{8}{c}{Global symmetry, $G_F$} \\[0.1cm]
         \hline
     & $SU(4)_{HC}$ & $SU(5)$ & $SU(3)_1$ & $SU(3)_2$  & $SU(3)^4$ & $U(1)_X$ & $U(1)_B$ & $U(1)_{A_1}$ & $U(1)_{A_2}$\\
     \hline
     $\psi$ & \textbf{6} & \textbf{5} & \textbf{1} & \textbf{1} & (\textbf{1}, \textbf{1}, \textbf{1}, \textbf{1}) & 0 & 0 & --18/5 & 0 \\[0.1cm]
     $\chi^i_u$ & \textbf{4} & \textbf{1} & \textbf{3} & \textbf{1} & (\textbf{3}, \textbf{1}, \textbf{1}, \textbf{1}) & --1/3 & --1/6 & 1 & 1 \\[0.1cm]
     $\tilde{\chi}^i_u$ & $\bar{\textbf{4}}$ & \textbf{1} & \textbf{1} & $\bar{\textbf{3}}$ & (\textbf{1}, $\bar{\textbf{3}}$, \textbf{1}, \textbf{1}) & 1/3 & 1/6 & 1 & 1 \\[0.1cm]
     $\chi^j_d$ & \textbf{4} & \textbf{1} & \textbf{3} & \textbf{1} & (\textbf{1}, \textbf{1}, \textbf{3}, \textbf{1}) & 1/6 & --1/6 & 1 & --1 \\[0.1cm]
     $\tilde{\chi}^j_d$ & $\bar{\textbf{4}}$ & \textbf{1} & \textbf{1} & $\bar{\textbf{3}}$ & (\textbf{1}, \textbf{1}, \textbf{1}, $\bar{\textbf{3}}$) & --1/6 & 1/6 & 1 & --1 \\[0.1cm]
     \hline 
     \hline
    \end{tabular}
    \end{adjustbox}
    \caption{Particle content and their charges in the model of Sec.~\ref{composite}. }
    \label{tab:UV_particles_compH}
\end{table}
 The fields of the strong sector and their  charge assignments are given in Table~\ref{tab:UV_particles_compH}. The field, $\psi$, transforms in the fundamental representation of the global group $SU(5)$ and as a 6-plet under the hypercolor gauge group. We assume that the hypercolor group becomes strongly coupled in the IR and this results in the $\psi$ forming a condensate that breaks $SU(5)\to SO(5)$; see Ref.~\cite{RSG_Comp_Higgs} for further assumptions regarding the running of the hypercolor group. This generates 14 pseudogoldstone bosons including the 4 corresponding to the Higgs doublet.
 
 The unbroken subgroup $SO(5) \times U(1)_X$ contains  the subgroup $SU(2)_L \times SU(2)_R$ which is spontaneously broken to its diagonal subgroup, the so-called  custodial group, upon electroweak symmetry breaking. The $SU(2)_L$ is gauged and identified with the SM  gauge group whereas only a linear combination of an abelian subgroup of $SU(2)_R$  and  $U(1)_X$ is gauged to obtain the SM hypercharge group; the hypercharge generator is given by, $Y = T_{3R} + X$, where $T_{3R}$ denotes the diagonal generator of $SU(2)_R$. 
 
As far as the SM fermion masses and mixings are concerned, Ref.~\cite{RSG_Comp_Higgs} utilises partial compositeness~\cite{kaplan, cp} that provides a  mechanism to generate the large flavor hierarchies of the SM.  Partial compositeness is realised by linearly coupling the SM fermions to composite baryonic operators  $\mathcal{O}_F$,
 \begin{align}
    \mathcal{L}_{\text{mix}} &= \frac{\lambda_{u_R}^{ij}}{\Lambda^{d_{U_R}-5/2}} \,u^{c,\,i} \,\mathcal{O}_{U_R}^j +\frac{\lambda_{u_L}^{ij}}{\Lambda^{d_{U_L}-5/2}}\, q^{i} \mathcal{O}_{U_L}^{c,\,j} \nonumber\\
    &+ \frac{\lambda_{d_R}^{ij}}{\Lambda^{d_{D_R}-5/2}} \,d^{c,\,i} \,\mathcal{O}_{D_R}^j + \frac{\lambda_{d_L}^{ij}}{\Lambda^{d_{D_L}-5/2}} \,q^{i} \,\mathcal{O}_{D_L}^{c,\,j} +  \text{{\rm h.c.}},
    \label{pcomp}
\end{align}
where $d_F$ is the conformal dimension of the operator $\mathcal{O}_F$;  $i, j = 1$-$3$, are flavor indices and the explicit form of the operators in terms of the fields in Table~\ref{tab:UV_particles_compH} is given by,
\begin{align}
    \mathcal{O}^{c}_{U_L,i}
    &\sim \epsilon_{ijk}\,\chi_u^j\,\psi\,\chi_u^k,
    &
    \mathcal{O}_{U_R}^{i}
    &\sim \epsilon^{ijk}\,
        \bar{\chi}_{u,j}\,\bar{\psi}\,\tilde{\chi}_{u,k},
    \nonumber\\
    \mathcal{O}^{c}_{D_L,i}
    &\sim \epsilon_{ijk}\,\chi_d^j\,\psi\,\chi_d^k,
    &
    \mathcal{O}_{D_R}^{i}
    &\sim \epsilon^{ijk}\,
        \bar{\chi}_{d,j}\,\bar{\psi}\,\tilde{\chi}_{d,k},
    \label{ops}
\end{align}
where the normalisation   above is  $4 \pi$ according to naive dimensional analysis(NDA)~\cite{gmanohar}. As in Ref.~\cite{RSG_Comp_Higgs}, to prohibit the mixing of the SM fermions to other possible baryonic operators, a $\mathbb{Z}_2$ symmetry---under which $\chi^i_{u,d}$ and the SM right-handed fermions are odd---is assumed. The   lagrangian in \eq{pcomp} explicitly breaks the global symmetry and plays a key role in generating the Higgs potential~\cite{AGASHE2005165, Panico_2016}.

When the strong sector confines, these operators become interpolating operators that generate composite 
states---the so-called partners of SM fermions---at the scale, $\mu\sim m_\star$. The lagrangian   terms involving these partner states then becomes,
 \begin{align}
 {\cal L}_{\rm partner} (\mu =m_\star) &=\frac{m_\star}{g_\star} \Bigg(
 {\lambda_{u_R}^{ij}}  u^{c,\,i} \,{U_R}^j + {\lambda_{u_L}^{ij}}  q^{i} {U_L}^{c,\,j}
    +  {\lambda_{d_R}^{ij}} d^{c,\,i} \,{D_R}^j + {\lambda_{d_L}^{ij}} q^{i} \,{D_L}^{c,\,j} +  \text{{\rm h.c.}} \Bigg).
    \label{pcomp2}
\end{align}
These states pair with  their conjugates  to form massive Dirac fermions that have a coupling  strength, $g_*$, with the Higgs. Integrating out these heavy Dirac fermions generates   the Yukawa interaction matrices for the SM fermions, 
\begin{equation}\label{leftright}
    Y_f^{ij}(m_\star) \sim \frac{\lambda^{ik}_{f_L}(m_\star)\,\lambda^{jk}_{f_R}(m_\star)}{g_*}.
\end{equation}
Following    Ref.~\cite{Redi} we assume, 
\bea
\lambda_{u_R, d_R}^{ij}\sim y_{u_R, d_R} e^{i \theta_R}\delta^{ij}, 
\label{thetar}
\eea
so that the final Yukawa matrices for the quarks become proportional to the left handed mixings, 
\bea\label{spurion}
Y_u^{ij} \sim  \frac{\lambda_{u_L}^{ij} y_{u_R}}{g_\star}\nn\\
Y_d^{ij} \sim \frac{\lambda_{d_L}^{ij} y_{d_R}}{g_\star}.
\eea
If we consider  $\lambda_{u_L}^{ij}$ and $\lambda_{d_L}^{ij}$ to be   spurions  that transform under the  $U(3)^3$ flavor symmetry of the SM quarks, their transformation properties are identical to those of SM Yukawa matrices. Therefore the couplings, $\lambda_{u_L}^{ij}$ and $\lambda_{d_L}^{ij}$, carry two independent phases, namely  the strong CP phase and a CKM-like phase.

In the  mass basis  we can write,
\bea
\lambda^{ij}_{u_L} &=& \mbox{diag}\{\hat{y}_{u_L},\, y_{c_L},\, y_{t_L}\}\nn\\
\lambda^{ij}_{d_L} &=& \mbox{diag}\{y_{d_L},\, y_{s_L},\, y_{b_L}\}
\eea
where, as in the SM, we perform independent flavor rotations of the up- and down-type components. The couplings $y_f$ evolve under RG  and  are given by~\cite{RSG_Comp_Higgs},
\begin{equation}\label{eq:RG_eq_yL}
    y_f (m_*) \simeq y_f(\Lambda) \left(\frac{m_*}{\Lambda}\right)^{d_f - 5/2}.
\end{equation}
at the IR scale, $m_{*} \sim 1$ TeV where we have neglected   sub-dominant RG corrections. Thus, even if the effective SM Yukawa couplings exhibit an anarchic structure in the UV with $\mathcal{O}(1)$ values, a large scale separation $\Lambda \gg m_*$ together with appropriate scaling dimensions, $d_F$, can generate the observed flavor hierarchies in the IR~\cite{Panico_2016}. For scales, $\mu \lesssim m_\star$, this theory also maps to the lagrangian in \eq{lint}.

Ref.~\cite{RSG_Comp_Higgs}    also considers some lagrangian  terms  involving the   $\psi$-field that explicitly break the global symmetry. The field  $\psi$  decomposes as, $\psi = (\Psi_+,\, \Psi_-,\,\psi_0)^T$, where, $\Psi_{\pm}$ are $SU(2)_L$ doublets with $T_{3R} = \pm 1/2$, while $\psi_0$ is a $SU(2)_L$ singlet. The doublets are further decomposed as $\Psi_{\pm} = (\psi_{\pm+},\,\psi_{\pm-})^T$ where the second index specifies $T_{3L} = \pm 1/2$. The additional terms considered in Ref.~\cite{RSG_Comp_Higgs} can be written as, 
\begin{align}\label{psil}
\mathcal{L}_{\psi}
&= m_0 \psi_0\psi_0
+ \frac{g_{4\psi}}{\Lambda_\psi^2}\,
\epsilon^{ab}
\left(\bar{\Psi}_{-,\,a}\,\bar{\sigma}_{\mu}\psi_0\right)
\left(\bar{\Psi}_{+,\,b}\,\bar{\sigma}^{\mu}\psi_0\right)
+ \mathrm{h.c.}
\\
&\supset m_0 \psi_0\psi_0
+ \frac{g_{4\psi}}{\Lambda_\psi^2}
\Big[
\left(\bar{\psi}_{-+}\,\bar{\sigma}_{\mu}\psi_0\right)
\left(\bar{\psi}_{+-}\,\bar{\sigma}^{\mu}\psi_0\right)
\Big]
+ \mathrm{h.c.}
\end{align}
Here, $m_0$ and $g_{4\psi}$ are taken to be real without loss of generality, as all the phases---including $\theta_R$ in \eq{thetar}---can be rotated to the couplings, $\lambda^{ij}_{u_L}$ and $\lambda^{ij}_{d_L}$, as shown in Ref.~\cite{RSG_Comp_Higgs}.

Finally we have an axion in our model with its usual coupling to the QCD gauge fields, 
\bea
{\cal L}_{a} &\supset&\frac{(\partial_\mu a)^2}{2} 
   +\frac{g_s^2}{16\pi^2} \frac{a}{f_a} \Tr (G_{\mu\nu}\widetilde G^{\mu\nu}).
\eea
In the limit the  first diagonal entry of $\lambda^{ij}_{u_L}$ in the mass basis, $\hat{y}_{u_L}$, vanishes we recover an exact non-anomalous symmetry; this is given  by the following transformations in the mass basis,
\begin{align}
    u^{c,i}
    &\longrightarrow e^{i\alpha\delta_{i1}}\,u^{c,i},
    &
    \frac{a}{f_a}
    &\longrightarrow \frac{a}{f_a}+\alpha,
    \nonumber\\
    \chi_u^i
    &\longrightarrow e^{iq_i\alpha}\,\chi_u^i,
    &
    \tilde{\chi}_{u,i}
    &\longrightarrow e^{-iq_i\alpha}\,\tilde{\chi}_{u,i},
    \label{global3}
\end{align}
where $\delta_{i1}$ in the first transformation is a Kronecker delta. Here,
\bea
    (q_1,q_2,q_3)
    =\left(-\frac12,\frac12,\frac12\right)
\eea
are the charges of the up-sector hyperfermions $\chi_u^i$.
The  contributions to the QCD and hypercolor
anomalies cancel pairwise, while the axion shift cancels
the QCD anomaly of the $u^{c,1}$ rotation.
 The parameter   $\hat{y}_{u_L}$  corresponds to the small parameter, $\hat{y}$, introduced in Sec.~\ref{conditions} and the axion mass vanishes in the limit this parameter vanishes. We now show that small instantons can additively generate a   contribution to $\hat{y}_{u_L}$ that is large enough to reproduce the correct up quark mass in the IR. 
\begin{figure}[t]
\hspace{1cm}
\begin{subfigure}{0.45\textwidth}
\hspace{-0.4cm}
\centering
\includegraphics[width=0.8\linewidth]{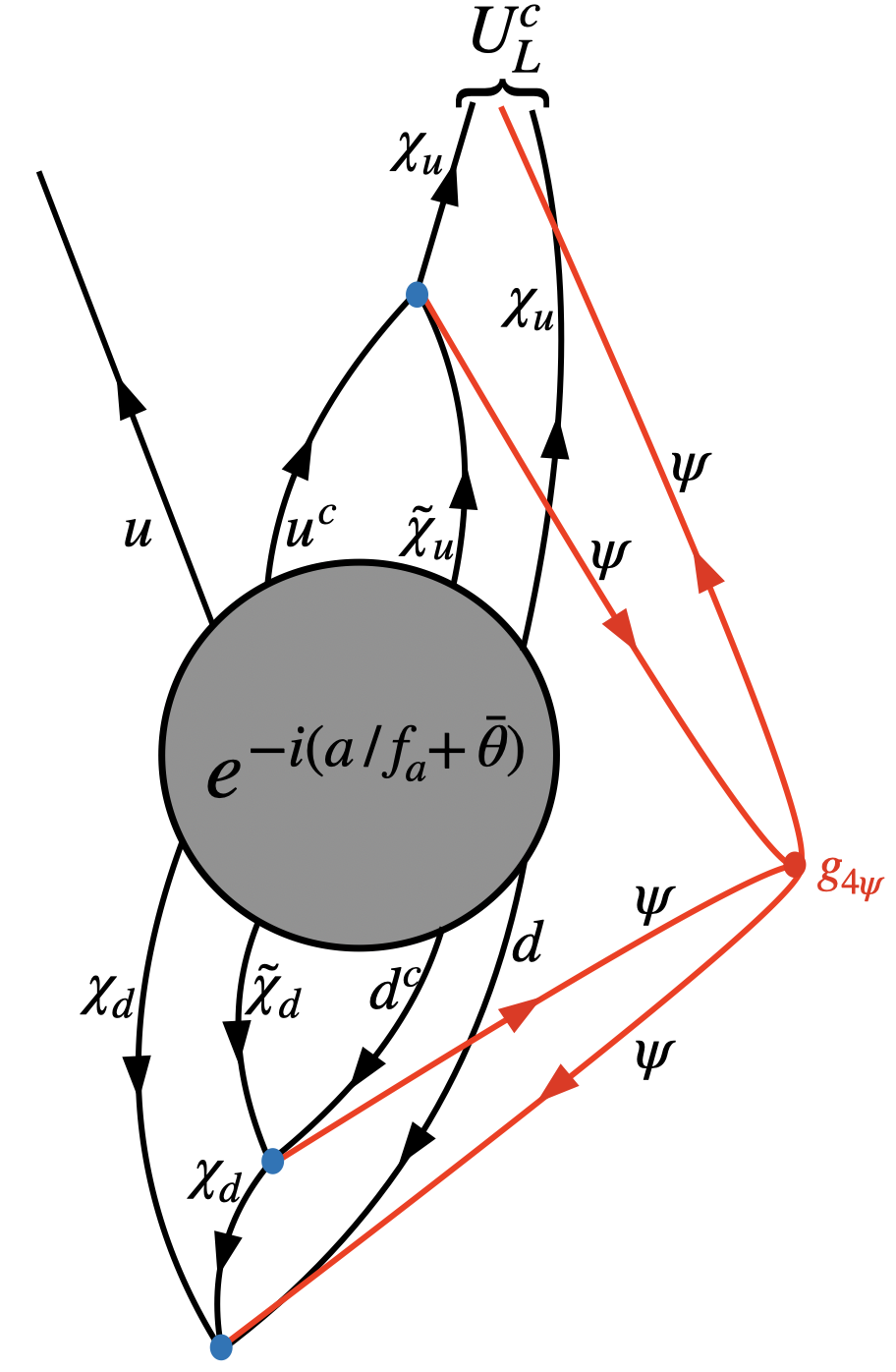}
\caption{}
\label{fig:up_quark_model2}
\end{subfigure}
\begin{subfigure}{0.45\textwidth}
\centering
\includegraphics[width=0.8\linewidth]{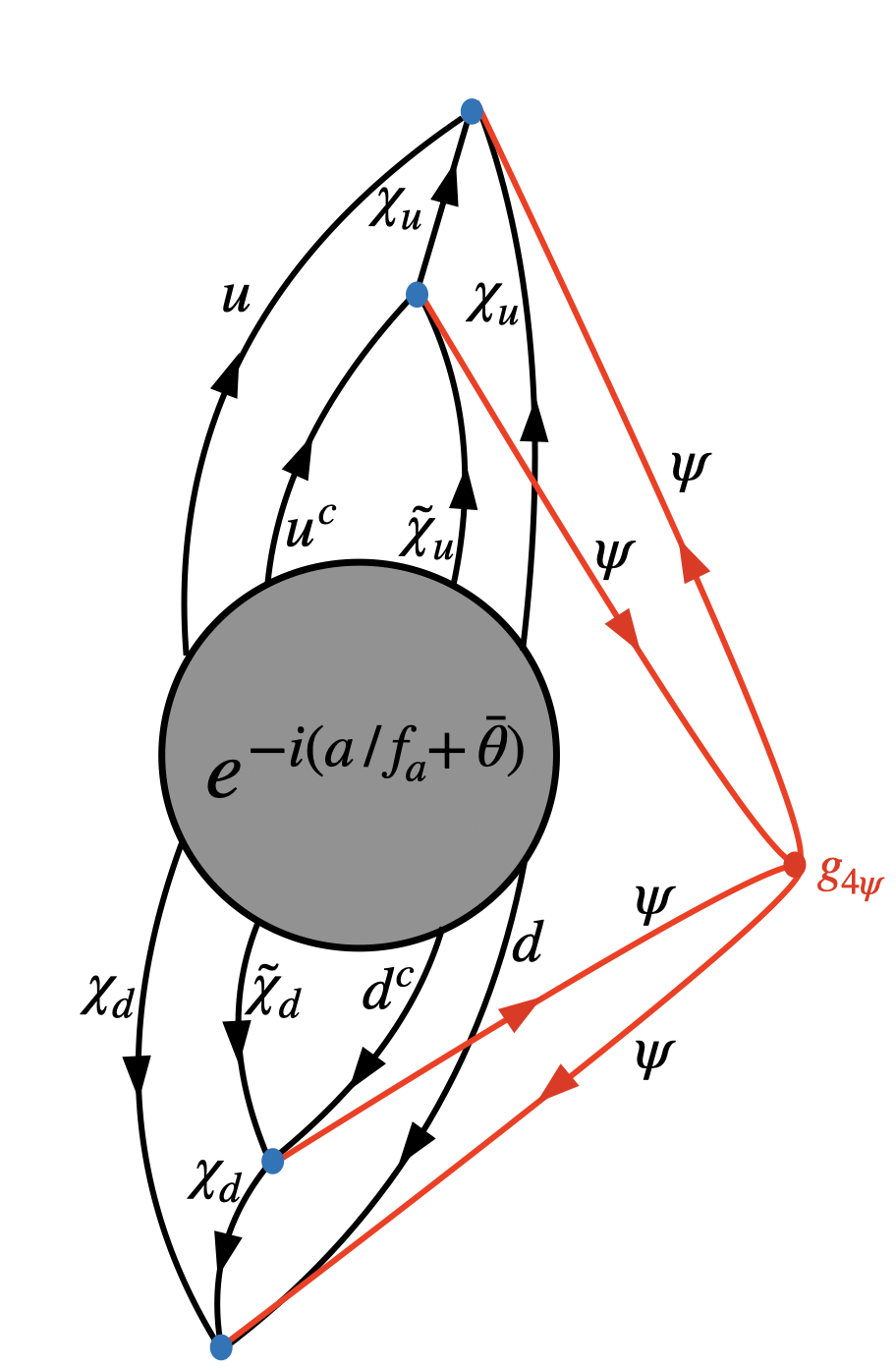} 
\caption{}
\label{fig:axion_potential_model2}
\end{subfigure}
\caption{Small instanton contributions to (a)  the left-handed up-quark mixing term, and (b) the axion potential in the model of Sec.~\ref{composite}. A diagram similar to the left one gives contribution to the left-handed down type mixing term. The above diagrams do not show the fermion lines corresponding to the second and third generations that can be closed in a similar manner using the elementary-composite mixing terms depicted by blue dots.}
\label{fig:composite_model}
\end{figure}

\subsection{Small Instanton Contributions}

Due to the large number, $N_F=30$, of   colored fermions in our model the QCD $\beta$-function changes sign~\cite{RSG_Comp_Higgs}, 
\begin{equation}
    \frac{d g_s}{d\ln\mu} = -\left(11 - \dfrac{2}{3}N_F \right) \frac{g^3_s}{16 \pi^2}
\end{equation}
and $g_s$   becomes strongly coupled at $M \sim 2000$ TeV.
This leads to sizeable small instanton effects at this scale that affect the axion mass calculation. Above this scale we assume that the growth of $\alpha_s$ is tamed by some  UV physics;  as in other models~\cite{Holdom1985, FlynnRandall1987, RSG_Comp_Higgs,Kitano:small_instantons} that consider the effect of small instantons on the axion mass, we keep this physics unspecified. 

\subsubsection*{Generating additive contribution to $\hat{y}_{u_L}$  from small instantons}
Small instantons generate a field-dependent  additive contribution to $\hat{y}_{u_L}$  that we call $\Delta y_{u_L}^{\rm SI}$. This contribution arises  from a diagram where   all the fermions   in the `t Hooft vertex are closed except   the $u$-quark and the fermions that constitute $\mathcal{O}^{c, \,1}_{U_L}$, see Fig.~\ref{fig:up_quark_model2}. From \eq{param}, we obtain the following parametric estimate for  the instanton-induced contribution to $y_{u_L}$ at the scale $M$, 
\begin{equation}\label{yukcomp}
    \frac{\Delta y_{u_L}^{\text{\rm SI}}}{4\pi} \sim   (16 \pi^2 \xi_{I}) \left(\dfrac{2\pi}{\alpha_s(M)}\right)^{6} e^{-\frac{2\pi}{\alpha_s(M)}}\left[\left(\frac{g_{4\psi}}{16\pi^2}\right)^3 \frac{{y}_{u_R}}{4\pi}\prod_{f=d,s,c,b,t} \frac{{y}_{f_L}}{4\pi} \frac{{y}_{f_R}}{4\pi}\right] e^{-i\left(\frac{a}{f_a}+\bar{\theta}\right)}
\end{equation}
where we have assumed $\Lambda_\psi \sim M$ in \eq{psil}. In the limit the couplings approach their non-perturbative values, the factor in the square bracket goes to 1. To obtain the above estimate we have  normalised the operators, following NDA, with the $4\pi$ factor mentioned below \eq{ops}. This implies that the blue vertices in Fig.~\ref{fig:up_quark_model2}  contribute a factor  $4 \pi y_{f_{L,R}}$,  $y_{f_{L,R}}$ being the appropriate elementary-composite mixing term.  Using $N_S=0$ and $N_F=30$ in \eq{xi} we  have for this case,  
\bea\label{xicomp}
\xi_I (30,0) = 0.106.
\eea
 We assume, ${y}_{f_L} (M)=y_{f_R}(M) = 4\pi$ for $f = s, c, b,t$. As far as the first generation singlet quarks are concerned, we take  ${y}_{u_R,d_R} (M)= 4\pi$ and  fix the anomalous dimensions of the  operators in \eq{ops} such that, $y_{u_R,d_R}(m_*)= {y}_{u_R,d_R}(M)/10$. For the first generation doublet quarks we fix the anomalous dimension for their partner operators, $d_{U^1_L,D^1_L}=7/2$ and fix ${y}_{u_L,d_L} (M)$ such that the measured SM fermion masses for the first generation can be reproduced using Eq.~\ref{eq:RG_eq_yL} and,
\bea \label{yucomp}
\hat{y}_u&\sim& \frac{\hat{y}_{u_L} y_{u_R}}{4\pi}\nn\\
  \delta y_{u}^{\rm SI}  &\sim& \dfrac{y^{\rm SI}_{u_L} \,y_{u_R}}{4\pi},
\eea
where we have assumed  $g_\star= 4 \pi$ in \eq{leftright}. For $\hat{y}_u \ll y_u$  we then obtain, 
\bea
    \frac{\delta y_u^{\rm SI}}{y_d} &\sim 16\pi^2 \xi_I^{(3)}\, \left(\dfrac{2\pi} {\alpha_s(M)}\right)^6\,e^{-\frac{2\pi}{\alpha_s(M)}} \left(\dfrac{g_{4\psi}}{16\pi^2}\right)^3.\label{eq:yuk_ratio_model2}
\eea
While \eq{yucomp} and \eq{eq:yuk_ratio_model2}  have been derived at the scale, $\mu \sim m_\star$, as in Sec.~\ref{product}, the ratio in \eq{eq:yuk_ratio_model2} is very stable under RG---in particular one can safely neglect the running between the scale $m_\star$ and the scale, $\mu \sim  2~{\rm GeV}$, in \eq{conditionratio}. We can therefore impose the condition in \eq{conditionratio} directly on the above ratio  to obtain,
\bea \label{pt3}
\left(\frac{2\pi} {\alpha_s(M)}\right)^6\,e^{-\frac{2\pi}{\alpha_s(M)}} \left(\frac{g_{4\psi}}{16\pi^2}\right)^3\sim 0.03
\eea
where we have used \eq{xicomp}. We see that this condition can be satisfied for large values, $\alpha_s(M)$, and an appropriate choice of $g_{4 \psi}$.

Finally, for the parameter choices and assumptions stated below \eq{xicomp}, the small instanton contribution to the down Yukawa is given by,
 \begin{equation}
   \delta \hat{y}_d^{\rm SI} \sim 16\pi^2 \xi_I^{(3)}\, \hat{y}_u \left(\dfrac{2\pi} {\alpha_s(M)}\right)^6\,e^{-\frac{2\pi}{\alpha_s(M)}} \left(\dfrac{g_{4\psi}}{16\pi^2}\right)^3.
 \end{equation}
\subsubsection*{Small instanton contribution to axion potential}
Small instantons also generate a potential for the axion when all the fermion lines of the `t Hooft vertex are closed as shown in Fig~\ref{fig:axion_potential_model2}. Using \eq{param} we obtain the following parametric expression for the potential, 
\begin{equation}\label{potcomp}
  V_{\rm SI}(a)\sim -  \xi_{I}^{(3)} \left(\dfrac{2\pi}{\alpha(M)}\right)^{6} e^{-\frac{2\pi}{\alpha(M)}}\left(\frac{g_{4\psi}}{16\pi^2}\right)^3\left[ \prod_{f = u, d, s, c, b, t} \frac{ {y}_{f_L}}{4\pi} \frac{ {y}_{f_R}}{4\pi}\right] M^4 \cos \left(\frac{a}{f_a}+\bar{\theta}\right),
\end{equation}
where  we have assumed $\Lambda_\psi \sim M$ in \eq{psil}. In the limit that the couplings approach their non-perturbative values, the factor in the square bracket goes to 1. Once again, using the normalisation  factor mentioned below \eq{ops}  results in a factor   $4 \pi y_{f_{L,R}}$ for each of the blue vertices in Fig.~\ref{fig:axion_potential_model2}. Using the assumptions and parameter values mentioned below \eq{xicomp} we obtain,
 \begin{equation} \label{eq:num_axionmass_model2}
    \hat{m}_{a,{\rm SI}}^2 \sim 1.7 \times 10^{-3} \,\xi^{(3)}_I\,\left(\frac{2\pi}{\alpha_s(M)}\right)^6 e^{-2\pi/\alpha_s(M)}  \left(\frac{g_{4\psi}}{16\pi^2}\right)^3 \,\hat{y}_{u_L} \frac{M^4}{f_a^2}.
\end{equation}
We   estimate this contribution using \eq{xicomp} and \eq{pt3} and then use \eq{axionmass} to obtain.
\begin{equation}\label{axionmasscomposite}
    m_{a}^2 \sim (9 \times 10^{12} \sqrt{ \hat{y}_{u_L}}  \;{\rm meV})^2 \left(\frac{10^9\;{\rm GeV}}{f_a}\right)^2 \left(\frac{M}{2000 \;\rm TeV}\right)^4\;.
\end{equation}
where the IR contribution in \eq{axionmass} is again negligible.

In Fig.~\ref{fig:massplot} we plot the mass of the axion in this model as a function of $\hat{y}=\hat{y}_{u_L}$. We see that the axion is lighter than the standard   expectation for, $\hat{y}_{u_L} \lesssim 10^{-25}$, and heavier otherwise. As in Sec.~\ref{product}, the smallness of $\hat{y}=\hat{y}_{u_L}$ may arise from a UV framework where \eq{global3} accidentally survives up to a certain operator dimension but is broken by higher dimensional Planck-suppressed operators.

While the contributions in \eq{yukcomp} and \eq{potcomp} give the leading small instanton contributions, one must to eliminate the possibility of  misaligned  sub-leading contributions to the axion potential that result in a residual strong-CP phase. These  were  considered  in detail in Ref.~\cite{RSG_Comp_Higgs}. One source of such a contribution is the fact  that, in addition to the strong-CP phase, the mixing terms,  $\lambda_{u_L}^{ij}$ and $\lambda_{d_L}^{ij}$, also contain a CKM-like phase. As in the SM any contribution involving this phase must include multiple  quark flavors and loops and is expected to be  suppressed. Finally  higher dimensional operators can also result in misaligned contributions that give a residual strong CP phase; as discussed in detail in Ref.~\cite{RSG_Comp_Higgs}, these effects can be within experimental bounds for a sufficiently high new physics scale.

\section{Cosmology and Phenomenology}
\label{pheno}

In this section we discuss the cosmological implications and the phenomenological constraints  for the models presented in Sec.~\ref{product} and Sec.~\ref{composite}. The QCD axion in these models can contribute to the relic density of dark matter via the misalignment mechanism, and in some regions of the parameter space  constitute the whole observed dark matter density. The calculation of the relic density is, however, altered because of the presence of UV contributions to the axion potential.  We first present the computation of the QCD axion contribution to the observed dark matter abundance and then discuss phenomenological bounds.

\paragraph{Contribution of axion dark matter to relic density}
We assume pre-inflationary breaking of Peccei-Quinn symmetry and a reheating temperature greater than the scale where small instantons become active, i.e.  $T_{\rm RH}\gg M$. As the universe expands and cools down to lower temperatures,  $T  \lesssim  M$, a potential is generated for the axion that is given by only the second term of \eq{fullpot}. The axion field is initially displaced from its minimum by an angle, $\Delta \theta_m$, the so-called misalignment angle.  At early times  the Hubble friction is high, i.e.  ${\cal H} \gg  m_a$, so that the axion field value and energy density remain approximately fixed to their initial values. The axion field starts oscillating at a temperature $T_{\rm osc}$ when $3{\cal H} \sim m_a$. For the range of $m_a$ considered, $T_{\rm osc}$, is always well below, $M$, so the temperature dependence of the small instanton contribution to the axion mass can be ignored. For temperatures below $\Lambda_{\rm QCD}$,  instanton effects again become active generating an IR  contribution to the axion potential; as discussed in Sec.~\ref{product} and~\ref{composite} this contribution, however, is negligible compared to the small instanton one and can be safely ignored in this analysis. 

Once the axion field starts oscillating  its energy density   falls off as $\rho_a \propto a^{-3}$, where $a$ is the scale factor. The relic abundance  of the axion, $\Omega_a h^2$, the ratio of the present-day axion energy density to the critical density, is   given by,
\begin{equation} 
\Omega_a h^2 \simeq 0.20\,(\Delta\theta_m)^2
\left(\frac{m_a}{5\,{\rm neV}}\right)^2
\left(\frac{f_a}{4\times10^{13}\,{\rm GeV}}\right)^2
\left(\frac{1 \,{\rm GeV}}{T_{\rm osc}}\right)^3.
\label{relic}
\end{equation}

\begin{figure}[t]
    \centering
    \includegraphics[width=1\linewidth]{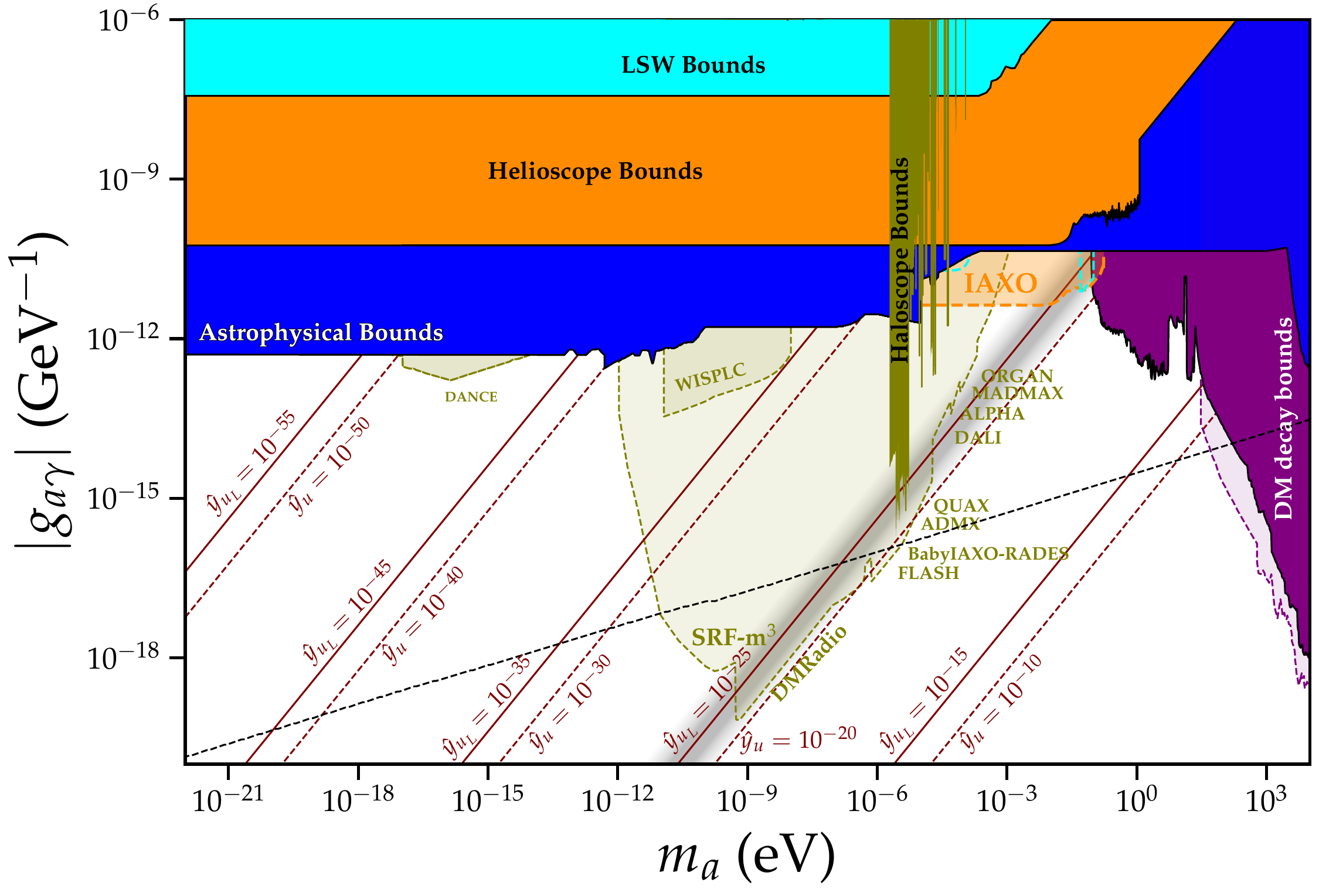}
    \caption{Constraints on the axion–photon coupling as a function of the axion mass. The dashed (solid) brown lines show the predictions of the model  of Sec.~\ref{product} (Sec.~\ref{composite}) for different values of the parameter   $\hat{y}_u$ ($\hat{y}_{u_L}$). We see that the QCD axion in these models can inhabit the region on both sides of the QCD axion band shown in gray. The axion in our models can fully (partly) account for the observed dark matter density below (above) the black dashed line. The experimental bounds have been obtained from Ref.~\cite{al} and references therein. The haloscope bounds from Ref.~\cite{al} have been rescaled according to the relic density in our model. The lightly shaded regions show the projected bounds from the experiments indicated by the corresponding colors. }
    \label{fig:Pheno_plot}
\end{figure}
\paragraph{Pion mass oscillations}

Due to the second term within the square-root sign in \eq{eq:QCD_potential} the axion field oscillations cause the pion mass to oscillate as well. This   in turn leads to oscillations of the nuclear charge radius, an effect that can be probed by optical atomic clocks according to the recent proposal of Ref.~\cite{gilad}. We see, however, that in the limit we are interested in, $\hat{y} \to 0$, these oscillations vanish so that the atomic clock probes of Ref.~\cite{gilad} will not be
sensitive to our model. 

\paragraph{Phenomenological constraints}
We now discuss the phenomenological constraints on the QCD axion in the models we have proposed. We focus here on the axion–photon coupling, as it offers a broad range of  powerful phenomenological probes, such as haloscopes, helioscopes, light-shining-through-wall experiments and astrophysical observations. In Fig.~\ref{fig:Pheno_plot} we have plotted the constraints from the current experiments in the $m_a-|g_{a\gamma}|$ parameter space where $g_{a \gamma}$ is the coefficient in the lagrangian to the term, $(a/4) \,F_{\mu\nu}\tilde{F}^{\mu\nu}$. We have from \eq{eq:aFF},
\bea
g_{a\gamma}=\frac{\alpha_{em}}{2\pi f_a}
\left(
\frac{E}{N}
-\frac{2}{3}\frac{4m_d + \hat{m}_u +4m^{\text{SI}}_u + \hat{m}_d^{\rm SI}}{m_d + \hat{m}_u + m^{\text{SI}}_u + \hat{m}_d^{\rm SI}}
\right)
\eea
In Fig.~\ref{fig:Pheno_plot}  we have set  $E/N=5/3$ so that the absolute value of the factor within the parenthesis above is $1$ for $\hat{m}_u \ll m_u$, a condition that holds in the relevant region of the plot;  this in turn implies that to a very good approximation,  $g_{a\gamma}= \alpha_{em}/ (2\pi f_a)$. We also show contours of constant $\hat{y}_u$  in the model of Sec.~\ref{product}  and  contours of constant $\hat{y}_{u_L}$  in the model of Sec.~\ref{composite}. We see that in both cases, for a given value of $g_{a \gamma}$,   the QCD axion can be both heavier and lighter than the standard value if we  suitably adjust  $\hat{y}_u$ or $\hat{y}_{u_L}$. Thus the axion in these models is  not  constrained to lie inside  the QCD  band and can inhabit the region  to its left as well as to its   right. 

The experimental bounds have been obtained from Ref.~\cite{al}. All the bounds apart from those arising from haloscopes are identical to  those in Ref.~\cite{al}. The haloscope bounds in~\cite{al}  assume that the axion constitutes whole of dark matter and therefore need to be modified in accordance with our relic density calculation. In Fig.~\ref{fig:Pheno_plot},  the region below the dashed black line shows the region where dark matter density can equal the relic density for some choice of $\Delta \theta_m \leq 1$; the misalignment angle is adjusted to give the observed relic abundance  in this region using \eq{relic}. In the region above the dashed line the dark matter density can only be a fraction of the observed dark matter density  for  $\Delta \theta_m \leq  1$; the misalignment angle in this region is set its maximal value, $\Delta \theta_m =  1$,  which in turn sets the   relic density in \eq{relic}. The haloscope bounds in the region above the dashed line   are  then rescaled to account for the reduced dark matter density. We find that, compared to the original constraints in~\cite{al}, this relaxes the bounds from   haloscope experiments  by a few orders of magnitude  in this region.

We see that the region, $|g_{a\gamma}|\lesssim 10^{-13}$, is mostly unconstrained.  A large part of this region will, however, be covered by future  experiments  such as  the IAXO helioscope~\cite{IAXO2019} and
haloscope programmes such as ORGAN~\cite{ORGAN2017},
MADMAX~\cite{MADMAX2019}, ALPHA~\cite{ALPHA2023},
DALI~\cite{DALI2024}, QUAX--$a\gamma$~\cite{QUAX2021},
ADMX~\cite{ADMX2018}, BabyIAXO--RADES~\cite{BabyIAXORADES2023},
FLASH~\cite{FLASH2023}, DMRadio~\cite{DMRadio2022},
SRF-$m^3$~\cite{SRF2020,SRF2022}
and WISPLC~\cite{WISPLC2022}.   We see that these experiments are particularly sensitive in the region to the left of the usual QCD axion band. Our work, therefore,    provides strong theory motivation for all these experiments.

 \section{Conclusions}

 In this work we have shown how the QCD axion can be both heavier and lighter than   the usual expectation, $m_a \sim m_\pi f_\pi/f_a$, in  the presence of small instantons. The main reason for this is that small instantons not only give  contributions to the axion potential but also    to the up-quark Yukawa. We show that if the latter   accounts for a large fraction of the up quark Yukawa the axion mass can be lighter than the usual expectation. 
 
 In Sec.~\ref{sec2} we present the  QCD axion mass computation in the presence of small instantons. For the small instanton contributions we consider a toy model and for the IR contribution below the QCD scale  we use the chiral lagrangian to derive a general expression in \eq{axionmass}. We  then discuss the conditions required to obtain a lighter than usual QCD axion. 

  We find that for the QCD axion to be lighter than the usual expectation   two conditions must be met: (1) a UV modification of QCD; and (2) a UV modification of the SM flavor sector. The first condition is essential to overcome  the non-perturbative suppression factor, $e^{- 2 \pi/\alpha_s(M)}$. The second is required   because there is a huge suppression factor in  the small instanton contribution to the up Yukawa coupling if the SM flavor structure is unaltered, see \eq{epsy}.

In Sec.~\ref{product} and Sec.~\ref{composite} we present two realistic models where these conditions can be realised. The first is a model where the QCD gauge group is a subgroup of a product of $SU(3)$ groups such that the gauge coupling is larger than the QCD coupling in the individual groups. Not all the SM fermions are charged under the same $SU(3)$ group which allows this model to overcome the flavor suppression of \eq{epsy}. The model of 
Sec.~\ref{composite} is a composite Higgs model where the QCD coupling becomes large in the UV as the number of colored constituent fermions is large. The anarchic flavor structure in the UV helps overcome the suppression of \eq{epsy}. In Fig.~\ref{fig:massplot} we show the  the QCD axion mass in these models can be both larger and smaller than the standard expectation.

 We show in Sec.~\ref{pheno} that the QCD axion in the above models  can explain part of---and in some regions the whole of---the observed dark matter abundance. The    axion  can account for the whole of dark matter in the region below the dashed line in Fig.~\ref{fig:Pheno_plot} that shows  phenomenological constraints in the $m_a-|g_{a\gamma}|$ plane. We find that contours of theory parameters can cover regions both to the left and right of the QCD  axion band. While the region below $|g_{a \gamma}| \lesssim 10^{-13}$ is currently unconstrained a whole array of proposed experiments   such as   IAXO ~\cite{IAXO2019}, ORGAN~\cite{ORGAN2017},
MADMAX~\cite{MADMAX2019}, ALPHA~\cite{ALPHA2023},
DALI~\cite{DALI2024}, QUAX--$a\gamma$~\cite{QUAX2021},
ADMX~\cite{ADMX2018}, BabyIAXO--RADES~\cite{BabyIAXORADES2023},
FLASH~\cite{FLASH2023}, DMRadio~\cite{DMRadio2022},
SRF-$m^3$~\cite{SRF2020, SRF2022}
and WISPLC~\cite{WISPLC2022}  are expected to probe this region in the coming years.  As the projected region these experiments can probe lies mostly to the left of the QCD axion band, our work provides a compelling theoretical motivation for these  programs.

\section*{Acknowledgements}
We  thank  Avik Banerjee and Gagandeep Singh for collaboration in the early stages of this work. We thank Abhishek Banerjee and  Marco Gorghetto for encouraging us to write up this idea. This work was carried out without any assistance from generative AI tools except for the use of GPT 6.1 just a few days before submission to identify typographical errors and possible inaccuracies in the equations. This did not lead to any significant changes. The authors take full responsibility for the final manuscript.  We acknowledge the support from the Department of Atomic Energy (DAE), Government of India, under Project Identification Number RTI 4012.

\appendix

\section{Small instanton contributions and their  parametric estimate}
\label{parametric}

In this appendix we discuss how to compute/estimate the small instanton contributions in the different models discussed in the main text. For the model in Sec.~\ref{product} the theory is well defined at high scales and we carry out  a computation of these effects in the dilute instanton gas approximation~\cite{tHooft_PRD}. In fact, these computations have been performed in the literature as we will soon review.  For the models presented in Sec.~\ref{toy} and Sec.~\ref{composite}, however, such a calculation is not possible, one of the reasons being that the physics above the scale, $M$, has not been completely specified. For these cases we  show how parametric estimates including factors of  $4\pi$ can be obtained  for any diagram involving  `t Hooft vertices.  For the case of small instanton contributions to the axion potential, our expression agrees with the estimate  carried out in Ref.~\cite{Csaki:Instanton_NDA}. 

\paragraph{Small instanton contributions in the model of Sec.~\ref{product}} In the dilute instanton gas approximation~\cite{tHooft_PRD}, the axion potential in an $SU(N_c)$ gauge theory is given by, 
\begin{equation}
    -\int d^4 x \,V(a) = W_{SU(N_c)} + \bar{W}_{SU(N_c)}
    \label{potW}
\end{equation}
where $(\bar{W}_{SU(N_c)})\,W_{SU(N_c)}$ is the vacuum-to-vacuum amplitude in one (anti-)instanton background. For a theory with $N_F$ colored Dirac fermions and $N_S$ colored complex scalars, the vacuum-to-vacuum transition amplitude in 1-instanton background is given by,
\begin{equation} \label{eq:path_int_vac}
     W_{SU(N_c)} \equiv \langle 0 | 0\rangle_{Q=1} = \frac{\int_{A^{\mu}_{\text{inst}.}} \mathcal{D}A_{\mu}\mathcal{D}\eta \mathcal{D}\bar{\eta}\mathcal{D} \Phi \mathcal{D}\Phi^{\dagger} \mathcal{D}\psi_f \mathcal{D}\bar{\psi}_f\, e^{-S_E}}{\int_{A^{\mu}_{cl}=0} \mathcal{D}A_{\mu}\mathcal{D}\eta \mathcal{D}\bar{\eta}\mathcal{D} \Phi \mathcal{D}\Phi^{\dagger} \mathcal{D}\psi_f \mathcal{D}\bar{\psi}_f\, e^{-S_E}},
\end{equation}
where the integral measure includes the gauge field $A_{\mu}$, the ghost fields $\eta$ and $\bar{\eta}$, the scalar fields $\Phi = (\phi_1, \dots, \phi_{N_S})^T$ and the fermion fields $\psi_f$, $f = 1,2,\dots,N_F$. The functional integration can be performed in semi-classical approximation, where the Euclidean action $S_E$ is expanded to second order in fields around the classical instanton solution,
\begin{equation}
    S_0 = \frac{8\pi^2}{g^2} + i\left(\frac{a}{f_a} + \bar{\theta}\right). 
\end{equation}
The computation of the path integral in an instanton background, to quadratic order in field fluctuations, was originally done by `t Hooft in Ref.~\cite{tHooft_PRD}. For a recent review of this computation in the context of the axion potential, see Refs.~\cite{Csaki:Instanton_NDA, Csaki:many_heavy_axion}. 

Instanton solutions are specified by a set of collective co-ordinates: the spacetime location of the instanton $x_0$, size $\rho$, and the orientation in the gauge group. The collective coordinates correspond to flat directions of the action and manifest as zero-modes in the path integral. In an instanton background, the fermions also develop zero-mode wavefunctions $\psi_f^{(0)}$, which are the eigenfunctions of the Dirac operator with zero eigenvalues, $-i\slashed{D}(\slashed{A}_{\text{inst.}})\psi_f^{(0)} = 0$. 
Upon separating out the zero modes and performing the functional integral over the non-zero modes, the vacuum-to-vacuum amplitude takes the form 
\begin{equation}\label{eq:vac_to_vac_amp}
    W_{SU(N_c)} = e^{-i(a/f_a + \bar{\theta})} \int d^4x_0 \int d\tilde{g} \int\frac{ d\rho}{\rho^5} D_{I}[\alpha(\rho^{-1})]  \int \mathcal{D}h \,e^{-S_0[h]}\int \prod_{f=1}^F \rho \,d\xi^{(0)}_f d\bar{\xi}^{(0)}_f e^{-S_{\text{int}}}.
\end{equation}
where $d \tilde{g}$ is the Haar measure over distinct gauge orientations of the instanton, normalized such that $\int d\tilde{g} = 1$; $d\xi_f^{(0)}, d\bar{\xi}_f^{(0)}$ are the Grassmann integration measures that project out the zero-mode fermion wave-function $\int d\xi^{(0)}\psi_f =  \psi^{(0)}_f$; $S_0[h]$ is the free action of the Higgs field $h$, and the instanton density $D_I[\alpha(\rho^{-1})]$ is given by 
\begin{equation}
    D_I[\alpha] = \xi_I (N_c,\,N_F,\,N_S) \left(\frac{2\pi}{\alpha}\right)^{2N_c} e^{-2\pi/\alpha}.
\end{equation}
Here,
\begin{equation}
    \xi_I (N_c,\,N_F,\,N_S) = \frac{C_1 \,e^{- C_3(N_S -2N_F)}}{(N_c-1)! (N_c-2)!} e^{-C_2 N_c} \,e^{-\frac{1}{12}(2N_F - N_S -2N_c)}
\end{equation}
in  the $\overline{\rm MS}$ scheme and the numerical coefficients, $C_1 = 0.466$, $C_2 = 1.678$ and $C_3 = 0.146$. For $N_c = 3$ the function $\xi_I$ reduces to,
\begin{equation} \label{eq:xi_Nc_3}
    \xi_I^{(3)} (N_F,\,N_S) = 2.5 \times 10^{-3} \,e^{0.0625\,(2N_F-N_S)}.
\end{equation}
To obtain the leading order non-vanishing contribution to the integral, one needs to expand the exponential $e^{-S_{\text{int}}}$ to terms containing all the $2N_F$ fermions, and all additional fields must be fully Wick contracted, which gives propagators. 

We now focus on the  action corresponding to that of the first site  of Sec.~\ref{product} and compute the contribution to the axion potential following Ref.~\cite{Csaki:many_heavy_axion},
\begin{equation}
    S_{\text{int}} = -\frac{i}{\sqrt{2}}\int d^4x  \left(\hat{Y}_u\, h(x)\,\bar{u}(x)u(x) + Y_d\, h(x)\,\bar{d}(x) d(x)\right). 
    \label{sint}
\end{equation}
Substituting this interaction term in Eq.~\ref{eq:vac_to_vac_amp}, and expanding to quadratic order gives the leading order contribution to $W_{SU(3)_1}$ as,
\begin{align}
    W_{SU(3)_1} = e^{-i(a/f_a + \bar{\theta})} \int d^4x_0 \int d\tilde{g} \int_{0}^{M^{-1}} \frac{ d\rho}{\rho^5} D_{I}[\alpha_{s_1}(\rho^{-1})]\, \mathcal{I}_F.
    \label{instint}
\end{align}
with the fermionic part given by,
\begin{equation}
    \mathcal{I}_F = -\frac{\hat{Y}_u Y_d}{2}\, \rho^2\int d^4x_1 \int d^4 x_2 \,\bar{u}^{(0)}(x_1) \,u^{(0)}(x_1) \,\bar{d}^{(0)}(x_2) \,d^{(0)}(x_2) \Delta_F(x_1-x_2)
    \label{IF}
\end{equation}
where $\Delta_F(x_1-x_2)$ is the free scalar propagator obtained by Wick contracting the Higgs field.  Here we have cut-off the $\rho$-integral at length-scales above, $\rho \sim 1/M$, as the bi-doublet field, $\Sigma_{12}$ gets a VEV; for a more careful treatment involving constrained instantons, see Ref.~\cite{Csaki:many_heavy_axion}. For the lower limit of integration, we assume that the model of Sec.~\ref{product} remains asymptotically free and can   be consistently extrapolated to arbitrarily small length-scales. A diagrammatic representation of  the above contribution  is given by  Fig.~\ref{fig:axion_potential_model1}.

As the  beta function for the $SU(3)_1$ group above the scale $M$ is known, we can explicitly calculate $\alpha_{s_1} (\rho^{-1})$. Using the one loop beta function we get, 
\bea
e^{- 2 \pi/\alpha_{s_1}(\rho^{-1})}=e^{- 2 \pi/\alpha_{s_1}(M)} (M \rho)^b
\eea
where we have taken $\mu=1/\rho$ and $b= \left(11- 2 N_F/3 - N_S/6\right)=55/6$ is the coefficient of the $\beta-$function. Carrying out the integral in \eq{IF} and neglecting the logarithmic $\rho$-dependence of the Yukawa couplings as well as that of $\alpha_{s_1}$ in the prefactor $(2 \pi/\alpha_{s_1}(M))^6$, we finally get the following expression for the axion potential from \eq{potW} and \eq{instint}, 
\bea
\label{massproduct}
V_{\rm SI}(a) = -\frac{2}{b - 4}\xi_I^{(3)}({2,\,3}) \left(\frac{2\pi}{\alpha_{s_1}}\right)^6 e^{-2\pi/\alpha_{s_1}} \,\,\frac{\hat{Y}_u Y_d}{24\pi^2} M^4\,\cos\left(\frac{a}{f_a} + \bar{\theta}\right).
\eea

We now turn to a computation of the small instanton contribution to the Yukawa coupling for the fermion, $u$, in \eq{sint}. We want to evaluate the additive instanton contribution, 
\begin{equation}\label{eq:eff_L_Yukawa}
    \Delta \mathcal{L}(x) =-\frac{h(x)}{\sqrt{2}}\left[\delta Y_u^{\text{SI}}(\rho)\,e^{-i(a/f_a + \bar{\theta})}\,\bar{u}_L(x) u_R(x)\, +h.c.\right]
\end{equation}
 To this end, we compute the matrix element,
\begin{equation}\label{eq:LSZ}
    \langle p',\,q|\,S\,| p\rangle = i\int d^4 x\,d^4x' d^4y \,e^{ip'.x'+iq.y-ip.x}(q^2 - m_{H}^2) \bar{\mathtt{u}}(p') \slashed{p}' \langle 0|T\{u(x') \bar{u}(x) h(y)\}|0\rangle \slashed{p} \,\mathtt{u}(p),
\end{equation}
where we have used the LSZ reduction formula to express the above matrix element in terms of the 3-point correlation function. Here, $\mathtt{u}(p)$ is the Dirac spinor corresponding to the fermion field $u$, and we have suppressed the spinor and color indices. For an explicit computation of the 3-point correlation function  in an instanton background we separate once again the bosonic and fermionic zero-modes and perform the functional integrals for the non-zero modes,
\begin{align}\label{eq:3-pt}
    G_3 \equiv\langle 0|T\{u(x') \bar{u}(x) h(y)\}|0\rangle &= e^{-i(a/f_a + \bar{\theta})} \int \frac{d^4 x_0 d\rho}{\rho^5} D_I[\alpha_{s_1}(\rho^{-1})] \int \mathcal{D}h\, e^{-S_0[h]} \nonumber\\
    &\times\int \prod_{f=u,\,d} \rho \,d\xi^{(0)}_f d\bar{\xi}^{(0)}_f \bar{u}(x) u(x') h(y) e^{-S_{\text{int}}}.
\end{align}
The leading order contribution to this integral is then given by
\begin{align}
    G_3 = \frac{Y_d}{\sqrt{2}} e^{-i(a/f_a + \bar{\theta})} &\int \frac{d^4 x_0 d\rho}{\rho^5} D_I[\alpha_{s_1}(\rho^{-1})] \,\rho^2 \int d^4 z \, u^{(0)}(x')\,\bar{u}^{(0)}(x)\,d^{(0)}(z)\,\bar{d}^{(0)}(z) \, \Delta_F(y-z).
\end{align}
 As the zero mode wave-function, $\psi^{(0)}_f$ is a function of $(x - x_0)$,  substituting the above expression into \eq{eq:LSZ}, and integrating over $x_0$ yields the momentum-conserving delta function from the exponential factors. We can then compute the 3-point function  and  use \eq{eq:LSZ} to obtain the matrix element. Matching this to the expected contribution from the term in \eq{eq:eff_L_Yukawa}, 
\begin{equation}\label{eq:matrix_elem}
    \langle p',\,q|\,S\,| p\rangle = -i (2\pi)^4 \delta^4 (p'+q-p) \,\delta Y_u^{\rm SI}\,e^{-i\left( a/f_a+\bar{\theta}\right)}\,\bar{\mathtt{u}}(p')\,\mathtt{u}(p)  
\end{equation}
 we obtain,
\bea \label{rhoint2}
\delta Y_u^{\rm SI} = -\frac{4\pi^2}{3}\xi_I^{(3)}(2,  3) \int_{0}^{M^{-1}} \frac{d\rho}{\rho} \left(\frac{2\pi}{\alpha_{s_1} (\rho^{-1})}\right)^{6} e^{-2\pi/\alpha_{s_1}(\rho^{-1})}\, Y_d.
\eea
This integration can be performed by a change in variable from $d(\ln\rho$) to $d(2\pi/\alpha_{s_1})$ using the RG equation,
\begin{equation}\label{eq:RG_alpha}
    d\left(\frac{2\pi}{\alpha_{s_1}}\right) = -b \,d(\ln\rho).
\end{equation}
This reproduces the result in Ref.~\cite{Agrawal:massless_up},
\begin{equation}
\label{upproduct}
    \frac{\delta Y_u^{\text{SI}}}{Y_d} = \frac{4\pi^2}{3b}\xi_I^{(3)}(2,  3)\,\Gamma\left(7, \,2\pi/\alpha_{s_1}(M)\right)
\end{equation}
where we ignore the variation of  $Y_d$. A diagrammatic representation of this contribution is given by  Fig.~\ref{fig:up_quark_model1}.

With the same approximations we can similarly compute the small-instantion contribution to the down-quark Yukawa. This gives, 
\bea \label{appdownsi}
    \delta \hat{Y}_d^{\text{SI}} = \frac{4\pi^2}{3b} \xi_I^{(3)}(2,\,3) \Gamma(7,\,2\pi/\alpha_{s_1}(M))\,\hat{Y}_u.
\eea

\paragraph{Parametric estimate of small instanton contributions} 
To calculate the small instanton contributions in Sec.~\ref{toy} and~\ref{composite} we will encounter integrals over the instanton size $\rho$ such as those in \eq{instint} and \eq{rhoint2}. To evaluate these integrals we therefore need to know the dependence of $\alpha_s$ in the UV which has been kept  unspecified in both Sec.~\ref{toy} and~\ref{composite}. We can nevertheless estimate these contributions by assuming that the dominant contribution to these effects arises from instantons of size, $\rho \sim 1/M$.  As far as the contributions shown in Fig.~\ref{fig:composite_model} are concerned, further technical complications are introduced due to  the presence of multiple loops.  We therefore derive a way to estimate these contributions parametrically including factors of $4\pi$.

The legs emanating from the `t Hooft vertex correspond to fermion zero-modes. These are then closed by the Yukawa interaction vertices and a Higgs propagator. For the purpose of $(4\pi)$-counting, we use the explicit expression for the fermion zero-modes given in~\cite{Csaki:Instanton_NDA}, which gives for each zero-mode $ \psi^{(0)}_f$ a factor $\frac{1}{ \pi}$. Each interaction vertex has an integral over spacetime $\int d^4 x \propto 2\pi^2 \int x^3 dx$, while each propagator comes with a momentum integral $\int \frac{d^4p}{(2\pi)^4} \propto \frac{1}{8\pi^2}\int p^3 dp$. So, to summarise, the $(4\pi)$-counting associated with the fermion zero-modes, spacetime integration arising from the interaction vertex and, propagators are given by,
\begin{equation}\label{eq:4pi_counting}
    \psi^{(0)}_f \sim \frac{1}{4\pi}, \quad \quad \quad \quad \int d^4 x \sim (4\pi)^2, \quad \quad \quad \quad\Delta_F \sim \frac{1}{(4\pi)^2}.
\end{equation}

This  motivates the following 
  expression to obtain a parametric estimate for any diagram involving a  `t Hooft vertex,
\begin{equation}\label{param}
     (4\pi)^{F_0} \times \frac{1}{(16\pi^2)^L}\, \times \xi_I (N_c,\,N_F,\,N_S)\,\left(\frac{2\pi}{\alpha(M)}\right)^{2N_c} e^{-2\pi/\alpha (M)} \times {\rm coupling~constants} \times M^{4-d}
\end{equation}
where $d$ is the dimension of the operator corresponding to the vertex and  we have assumed that dominant contributions arise from instantons of size, $\rho \sim 1/M$, $M$ being the scale where the gauge coupling reaches its maximal value. Here, $F_0$ is the number of fermionic zero-modes projecting out of the `t Hooft vertex and $L$ is the number of loops in the diagram. For a connected diagram, $L = I-V+1$, where $I$ is the number of internal lines and $V$ is the number of vertices including the `t Hooft vertex.

We now explain the central idea behind the parametric expression in \eq{param}. Note that the $(4 \pi)$-counting in \eq{param} matches that of a standard Feynman diagram apart from the overall $(4 \pi)^{F_0}$ factor.  Consider first the zero modes that become external lines in a 't Hooft diagram. For these zero modes there is, in the path-integral computation, a factor of $1/(4 \pi)$ from  \eq{eq:4pi_counting} and a factor of $(4 \pi)^2$ from a $\int d^4 x$ integral in the LSZ reduction formula---thus giving   an overall $4 \pi$ factor.  For fermionic zero modes that become internal lines in a 't Hooft diagram, each line carries a factor $1/(4 \pi)$ from  \eq{eq:4pi_counting} and not the usual $1/(4 \pi)^2$ factor carried by internal fermionic propagators in a Feynman diagram. Thus  in the path-integral calculation each internal zero-mode   line gives us a factor of $4 \pi$ relative to a standard Feynman diagram computation.  This explains the $(4 \pi)^{F_0}$ factor. Note that,  as far as  parametric estimates of  contributions to the axion potential are concerned, this counting rule matches that   in Ref.~\cite{Csaki:Instanton_NDA}. If we use the expression in \eq{param} to estimate the small instanton contribution to the up Yukawa and axion potential in Sec.~\ref{product}, the final expression for the axion mass we obtain differs from  \eq{axionmassproduct}---which uses the more detailed computation described in the first part of this appendix--- by only an ${\cal O} (1)$ factor. 

Note that in both the calculation presented in the first part of this appendix and the parametric estimate above, there are possible small instanton contributions  that we are likely missing. For instance it should be possible to close the fermion legs in Fig.~\ref{fig:axion_potential_model1} by the couplings, $(\delta y_u^{\rm SI} e^{-i (a/f_a + \bar{\theta})})^\star$ and   $(\delta \hat{y}_d^{\rm SI} e^{-i (a/f_a + \bar{\theta})})^\star$, thus resulting in a cosine term. These are, however, multi-instanton contributions that are higher order within the dilute instanton gas approximation that we are assuming.

\section{Axion mass in the presence of small-instantons in the chiral lagrangian with 3-flavors}\label{app:B}
In this appendix we compute the axion mass in the presence of small instantons using an effective low energy description of QCD with three light flavors, i.e.  $F=3$. As in Sec.~\ref{chiral} we will assume, $\hat{m}_{a,{\rm SI}}^2\ll \Lambda^2_{\rm QCD}$. The form of the low-energy lagrangian is same as in \eq{lchi}, 
\begin{align}
    \mathcal{L}^{3F}_{\chi} &= \frac{f_{\pi}^2}{4}\mbox{ Tr }\partial_{\mu}U\partial^{\mu}U^{\dagger} + B_0f_{\pi}^3 \mbox{ Tr}(M_QU + M_Q^{\dagger}U^{\dagger}) - F(e^{-i\left(\frac{ a}{f_a} +\bar{\theta}\right)} \mbox{ det}\,U)  \nonumber\\
    &+  \hat{m}_{a,{\rm SI}}^2 f_a^2 \text{ cos}\left(\frac{a}{f_a} + \bar{\theta}\right), \label{lchi3}
\end{align}
but in this case we define $U = \text{exp}(i\Pi)$ as,
\begin{equation}
    \Pi = 2\pi^aT^a/f_{\pi} + \eta'\mathbb{I}_3/f_{\eta'}   = \frac{1}{f_{\pi}}\begin{pmatrix}
        \pi^0 +\frac{1}{\sqrt{3}}\eta & \sqrt{2}\pi^+ & \sqrt{2}K^+ \\
        \sqrt{2}\pi^- & -\pi^0 + \frac{1}{\sqrt{3}}\eta & \sqrt{2}K^0 \\
        \sqrt{2} K^- & \sqrt{2} \bar{K}^0 & -\frac{2}{\sqrt{3}}\eta
    \end{pmatrix} + \frac{\eta'}{f_{\eta'}}\mathbb{I}_3.
\end{equation}
Here the 8 pseudogoldstone bosons are given by the three pions ($\pi^{\pm}, \,\pi^0$), four kaons ($K^{\pm},\, K^0,\, \bar{K}^0$) and $\eta$, while $\eta'$ is associated with the breaking of $U(1)_A$. In \eq{lchi3} the quark mass matrix is given by,
\begin{equation}
     M_Q = \text{diag}\{\hat{m}_u + m_u^{\rm SI}\,e^{-i(a/f_a + \bar{\theta})},\;m_d + \hat{m}_d^{\rm SI}\,e^{-i(a/f_a + \bar{\theta})},\;m_s\}.
\end{equation}
The parameters, $m_d$ and $m_s$, have been chosen to be real without loss of generality.

As discussed in Sec.~\ref{chiral} the dominant contribution to the $\eta'$-potential arises from the chiral anomaly which gives a contribution of the form, 
\begin{equation}
   V({\eta'})\sim m^4 \left(\bar{\theta} + \frac{a}{f_a} - F\frac{\eta'}{f_{\eta'}}\right)^2
\end{equation}
where $F$ is the number of flavors. The equation of motion for $\eta'$ yields at low energies, 
\begin{equation}
  \eta' = \frac{f_{\eta'}}{3}\left(\frac{a}{f_a} + \bar{\theta}\right). 
\end{equation}
Integrating out $\eta'$ we get for the IR contribution to the axion potential,
\begin{align}
   V^{\rm (IR)}(a) = -2B_0 f_{\pi}^3 &\Bigg[\hat{m}_u\, \text{cos}\left(\frac{a/f_a+\bar{\theta}}{3} + \frac{\pi^0}{f_{\pi}} + \frac{\eta}{\sqrt{3}f_{\pi}}\right) + m_u^{\text{\rm SI}}\, \text{cos}\left(\frac{2(a/f_a+\bar{\theta})}{3} - \frac{\pi^0}{f_{\pi}} -\frac{\eta}{\sqrt{3}f_{\pi}}\right) \nonumber\\
    &+ m_d\, \text{cos}\left(\frac{a/f_a+\bar{\theta}}{3} - \frac{\pi^0}{f_{\pi}} + \frac{\eta}{\sqrt{3}f_{\pi}}\right) + \hat{m}_d^{\rm SI} {\rm cos}\left(\frac{2(a/f_a + \bar{\theta})}{3} + \frac{\pi^0}{f_{\pi}} - \frac{\eta}{\sqrt{3}f_{\pi}}\right)\nonumber\\
    &+ m_s \text{cos}\left(\frac{a/f_a+\bar{\theta}}{3} - \frac{2\eta}{\sqrt{3}f_{\pi}}\right) \Bigg].
\label{eq:axion_potential_3F}
\end{align}

The next step is to integrate out the pions and the $\eta$. To linear order in the fields, the equation of motion for the pions and $\eta$ give at low energies,
\begin{align}
    \begin{pmatrix}
        \hat{m}_u + m_u^{\text{\rm SI}} + m_d + \hat{m}_d^{\rm SI}& \frac{1}{\sqrt{3}}(\hat{m}_u + m_u^{\text{\rm SI}} - m_d - \hat{m}_d^{\rm SI}) \\
         \hat{m}_u + m_u^{\text{\rm SI}} - m_d - \hat{m}_d^{\rm SI} & \frac{1}{\sqrt{3}}(\hat{m}_u + m_u^{\text{\rm SI}} + m_d + \hat{m}_d^{\rm SI} + 4m_s) 
    \end{pmatrix}
    \begin{pmatrix}
        \pi^0 \\
        \eta
    \end{pmatrix} \nonumber\\
    = -\frac{a/f_a}{3} \begin{pmatrix}
         \hat{m_u} - m_d - 2m_u^{\text{\rm SI}} + 2\hat{m}_d^{\rm SI} \\
         \hat{m_u} + m_d - 2m_s - 2m_u^{\text{\rm SI}} - 2\hat{m}_d^{\rm SI}
    \end{pmatrix}.
\end{align}
Substituting this in \eq{eq:axion_potential_3F}, and expanding to quadratic order in $a$ gives for the axion mass,
\begin{equation}\label{3flvaxion}
  m_a^2 = 2B \frac{f_{\pi}^3}{f_a^2} \frac{m_s(\hat{m}_u +\hat{m}_d^{\rm SI})(m_d+m_u^{\rm SI}) + \hat{m}_u m_d(m_u^{\rm SI} + \hat{m}_d^{\rm SI}) +m_u^{\rm SI}\hat{m}_d^{\rm SI}(\hat{m}_u +m_d)}{m_s(\hat{m}_u + m_u^{\rm SI} + m_d + \hat{m}_d^{\rm SI}) + (\hat{m}_u + m_u^{\rm SI})(m_d + \hat{m}_d^{\rm SI})} + \hat{m}_{a,{\rm SI}}^2.
\end{equation}
In all our models, as the small parameter $\hat{y} \to 0$, we get, $\hat{m_u} \to 0$, $\hat{m}_d^{\rm SI} \to 0$ and $\hat{m}_{a,{\rm SI}}^2 \to 0$; the axion mass in \eq{3flvaxion} vanishes in this limit. The  expression  in \eq{3flvaxion} also reduces to the two-flavor  expression in \eq{axionmass} if we ignore ${\cal O}(m_d/m_s)$ terms. 

\section{Flavor corrections and the residual strong CP phase in the model with modified gauge groups}
\label{app:flavorCP}

In this appendix, we discuss how higher order effects can lead to a
small residual strong CP phase once higher order effects are included
in the model of Sec.~\ref{product}. We show that this residual phase is expected to be within experimental bounds.  Going beyond the leading effects
discussed in Sec.~\ref{product}, one way to parametrise higher order effects to the different
terms in the low energy lagrangian in \eq{lint} is
as follows,
\bea
{\cal L} &\supset&
\frac{g_s^2}{16\pi^2}
\left(\frac{\phi}{f_a}+\delta_1\right)
\text{Tr}\,G_{\mu\nu}\widetilde G^{\mu\nu}
+\hat{Y}_u\Lambda_{\rm SI}^4
\left[
\cos\frac{\phi}{f_a}
+\varepsilon\cos\left(\frac{\phi}{f_a}+\delta_2\right)
\right]
\nonumber\\
&&-\left[
y_u(\phi)\frac{h}{\sqrt{2}}u u^c
+y_d(\phi)\frac{h}{\sqrt{2}}d d^c
+\text{{\rm h.c.}}
\right].
\eea
where we have defined $\phi/f_a\equiv a/f_a+\theta_1$ and,
\bea
y_u(\phi)
&=&\hat y_u
+\delta y_u^{\rm SI_1}e^{-i\phi/f_a}
+\delta y_u^{\rm SI_2}e^{-i(\phi/f_a+\delta_3)},
\\
y_d(\phi)
&=&y_d+\hat y_u F_d(\phi/f_a).
\eea
Notice that as $\hat{y}_u \to 0$   we recover the symmetry in \eq{global2} and hence all the contributions to the axion potential are proportional to this parameter.  Here, $\hat{y}_uF_d(\phi/f_a)$ is a dimensionless function containing  instanton
and flavor factors and includes the correction, $\delta \hat{y}_d^{\rm SI}$, discussed in the main text.
Without any loss of generality $\hat y_u$, $y_d$ and the remaining
quark masses can be chosen to be real, and we can take
$\delta y_u^{\rm SI_{1,2}}\geq0$. In this convention the down-quark
mass phase that survives at $\hat y_u=0$ is absorbed into $\delta_1$.
The remaining down-quark phase is therefore given by,
\bea
\arg[y_d(\phi)]
\sim
\frac{\hat y_u}{|y_d (\phi)|}
\operatorname{Im}F_d(\phi/f_a)= \frac{\hat y_u}{|y_d (\phi)|} \sin \delta_4 (\phi) .
\label{eq:cp_down_phase}
\eea

The field dependent strong CP phase is given by,
\bea
\bar\theta_{\rm QCD}(\phi)
&=&\frac{\phi}{f_a}+\delta_1
+\arg[y_u(\phi)]+\arg[y_d(\phi)].
\label{phitheta}
\eea
To obtain the final value of the strong CP phase we need to evaluate
\eq{phitheta} at the vacuum expectation value of the axion field,
$\bar\theta_{\rm QCD}
=\bar\theta_{\rm QCD}(\langle\phi\rangle)$.
As the IR contribution to the potential is negligible compared to
the small instanton one, we ignore the former while minimizing the
axion potential to obtain,
\bea
\langle\phi\rangle/f_a\sim-\varepsilon\sin\delta_2.
\eea
Substituting in \eq{phitheta} we get, 
\bea
\bar\theta_{\rm QCD}
&\sim&
\delta_1-r_2\sin\delta_3
-r_u\varepsilon\sin\delta_2+
r_u \frac{|m_u|}{|m_d|} \sin \delta_4
\label{eq:cp_theta_parameters}
\eea
where,
\begin{align}
r_1&=\frac{v\,\delta y_u^{\rm SI,1}}{\sqrt{2}|m_u|},
&
r_2&=\frac{v\,\delta y_u^{\rm SI,2}}{\sqrt{2}|m_u|},
&
r_u&=\frac{v\,\hat y_u}{\sqrt{2}|m_u|}.
\label{eq:cp_mass_ratios}
\end{align}
Here $|m_u|=v|y_u(\langle\phi\rangle)|/\sqrt{2}$, $|m_d|=v|y_d(\langle\phi\rangle)|/\sqrt{2}$, and all masses
and Yukawa couplings are evaluated at the same scale.

In the limit $r_u\to0$, \eq{eq:cp_theta_parameters} reduces to 
the value estimated in Ref.~\cite{Agrawal:massless_up},
\begin{equation}
\left.\left(\delta_1-r_2\sin\delta_3\right)\right|_{r_u\to 0}
=\Delta\bar\theta_{\rm AH}\lesssim10^{-10}.
\label{eq:cp_AH_matching}
\end{equation}
In our model the estimate in \eq{eq:cp_AH_matching} will get
further corrections due to the terms proportional to $r_u$ in
\eq{eq:cp_theta_parameters},  
and possible corrections to $\left(\delta_1-r_2\sin\delta_3\right)$
that arise for $r_u\neq0$. For most of the range shown in Fig~\ref{fig:massplot} and~\ref{fig:Pheno_plot} we have $r_u<10^{-10}$ which ensures that
these corrections are below ${\cal O}(10^{-10})$, provided the
coefficients multiplying $r_u$ are at most of ${\cal O}(1)$. The coefficients multiplying $r_u$ are, in fact,  likely to be much smaller than  ${\cal O}(1)$ due to loop  and 
flavor suppression factors which would allow much larger values of $r_u$.

Finally let us consider the effect of   small non-vanishing Yukawa coupling for the strange and bottom quark, $\bar{y}_{s}$ and $\bar{y}_{b}$,  in the deep UV, $\mu \gg M$. We call the  $\theta$-angle for the second and third site---that can no longer  be rotated away---$\theta_2$ and $\theta_3$, respectively.   We choose $\bar{y}_{s}$ and $\bar{y}_{b}$ to be real.  Including   the leading small instanton contribution    the full Yukawa coupling of the strange and bottom quarks develop a phase, 
\bea
y_{s,b} 
&=&  \bar{y}_{s,b}
+\delta y_{s,b}^{\rm SI_{1}}e^{-i \theta_{2,3}}.
\eea
We see that the argument of the above couplings do not cancel $\theta_{2,3}$. The difference,  can be rotated into $\delta_1$. We see that as long as, 
\bea
 \bar{y}_{s,b}/\delta y_{s,b}^{\rm {SI,1}}\ll 10^{-10}
\eea
this will not result in an unacceptably large contribution to the final strong CP phase. 

\bibliographystyle{JHEP}
\bibliography{References}

@article{Wess:1971yu,
    author  = "Wess, J. and Zumino, B.",
    title   = "{Consequences of anomalous Ward identities}",
    journal = "Phys. Lett. B",
    volume  = "37",
    pages   = "95--97",
    year    = "1971",
    doi     = "10.1016/0370-2693(71)90582-X"
}

@article{koren,
    author = "Cordova, Clay and Hong, Sungwoo and Koren, Seth",
    title = "{Noninvertible Peccei-Quinn Symmetry and the Massless Quark Solution to the Strong CP Problem}",
    eprint = "2402.12453",
    archivePrefix = "arXiv",
    primaryClass = "hep-ph",
    doi = "10.1103/PhysRevX.15.031011",
    journal = "Phys. Rev. X",
    volume = "15",
    number = "3",
    pages = "031011",
    year = "2025"
}

@article{banerjeehook,
    author = "Banerjee, Abhishek and Buen-Abad, Manuel A. and Hook, Anson",
    title = "{Constructing a light QCD axion without $1/N$ tuning}",
    eprint = "2507.02049",
    archivePrefix = "arXiv",
    primaryClass = "hep-ph",
    doi = "10.1103/xdyq-b67n",
    journal = "Phys. Rev. D",
    volume = "112",
    number = "7",
    pages = "075027",
    year = "2025"
}

@article{gilad,
    author = "Banerjee, Abhishek and Budker, Dmitry and Filzinger, Melina and Huntemann, Nils and Paz, Gil and Perez, Gilad and Porsev, Sergey and Safronova, Marianna",
    title = "{Oscillating Nuclear Charge Radii as Sensors for Ultralight Dark Matter}",
    eprint = "2301.10784",
    archivePrefix = "arXiv",
    primaryClass = "hep-ph",
    reportNumber = "WSU-HEP-2301",
    doi = "10.1103/37vw-gc1r",
    journal = "Phys. Rev. Lett.",
    volume = "135",
    number = "22",
    pages = "22",
    year = "2025"
}

@article{Witten:1983tw,
    author  = "Witten, Edward",
    title   = "{Global Aspects of Current Algebra}",
    journal = "Nucl. Phys. B",
    volume  = "223",
    pages   = "422--432",
    year    = "1983",
    doi     = "10.1016/0550-3213(83)90063-9"
}

@article{witten,
title = {Large N chiral dynamics},
journal = {Annals of Physics},
volume = {128},
number = {2},
pages = {363-375},
year = {1980},
issn = {0003-4916},
doi = {https://doi.org/10.1016/0003-4916(80)90325-5},
url = {https://www.sciencedirect.com/science/article/pii/0003491680903255},
author = {E Witten}
}

@article{Agrawal:massless_up,
    author = "Agrawal, Prateek and Howe, Kiel",
    title = "{A Flavorful Factoring of the Strong CP Problem}",
    eprint = "1712.05803",
    archivePrefix = "arXiv",
    primaryClass = "hep-ph",
    reportNumber = "FERMILAB-PUB-17-562-PPD",
    doi = "10.1007/JHEP12(2018)035",
    journal = "JHEP",
    volume = "12",
    pages = "035",
    year = "2018"
}

@article{Agrawal:many_heavy_axion,
    author = "Agrawal, Prateek and Howe, Kiel",
    title = "{Factoring the Strong CP Problem}",
    eprint = "1710.04213",
    archivePrefix = "arXiv",
    primaryClass = "hep-ph",
    reportNumber = "FERMILAB-PUB-17-500-T",
    doi = "10.1007/JHEP12(2018)029",
    journal = "JHEP",
    volume = "12",
    pages = "029",
    year = "2018"
}

@article{Csaki:Instanton_NDA,
    author = "Cs{\'a}ki, Csaba and D'Agnolo, Raffaele Tito and Kuflik, Eric and Ruhdorfer, Maximilian",
    title = "{Instanton NDA and applications to axion models}",
    eprint = "2311.09285",
    archivePrefix = "arXiv",
    primaryClass = "hep-ph",
    doi = "10.1007/JHEP04(2024)074",
    journal = "JHEP",
    volume = "04",
    pages = "074",
    year = "2024"
}

@article{HoldomPeskin1982,
    author  = "Holdom, Bob and Peskin, Michael E.",
    title   = "{Raising the Axion Mass}",
    journal = "Nucl. Phys. B",
    volume  = "208",
    pages   = "397--412",
    year    = "1982",
    doi     = "10.1016/0550-3213(82)90228-0"
}

@article{DiLuzio:2016sbl,
    author        = "Di Luzio, Luca and Mescia, Federico and Nardi, Enrico",
    title         = "{Redefining the Axion Window}",
    journal       = "Phys. Rev. Lett.",
    volume        = "118",
    pages         = "031801",
    year          = "2017",
    doi           = "10.1103/PhysRevLett.118.031801",
    eprint        = "1610.07593",
    archivePrefix = "arXiv",
    primaryClass  = "hep-ph"
}

@article{Farina:photophilic,
    author        = "Farina, Marco and Pappadopulo, Duccio
                     and Rompineve, Fabrizio and Tesi, Andrea",
    title         = "{The photo-philic QCD axion}",
    journal       = "JHEP",
    volume        = "01",
    pages         = "095",
    year          = "2017",
    doi           = "10.1007/JHEP01(2017)095",
    eprint        = "1611.09855",
    archivePrefix = "arXiv",
    primaryClass  = "hep-ph"
}

@article{Agrawal:2017cmd,
    author        = "Agrawal, Prateek and Fan, JiJi
                     and Reece, Matthew and Wang, Lian-Tao",
    title         = "{Experimental Targets for Photon Couplings
                     of the QCD Axion}",
    journal       = "JHEP",
    volume        = "02",
    pages         = "006",
    year          = "2018",
    doi           = "10.1007/JHEP02(2018)006",
    eprint        = "1709.06085",
    archivePrefix = "arXiv",
    primaryClass  = "hep-ph"
}

@article{Rubakov:heavy_axion,
    author        = "Rubakov, V. A.",
    title         = "{Grand unification and heavy axion}",
    journal       = "JETP Lett.",
    volume        = "65",
    pages         = "621--624",
    year          = "1997",
    doi           = "10.1134/1.567390",
    eprint        = "hep-ph/9703409",
    archivePrefix = "arXiv"
}

@article{Fukuda:visible_axion,
    author        = "Fukuda, Hajime and Harigaya, Keisuke
                     and Ibe, Masahiro and Yanagida, Tsutomu T.",
    title         = "{Model of visible QCD axion}",
    journal       = "Phys. Rev. D",
    volume        = "92",
    pages         = "015021",
    year          = "2015",
    doi           = "10.1103/PhysRevD.92.015021",
    eprint        = "1504.06084",
    archivePrefix = "arXiv",
    primaryClass  = "hep-ph"
}

@article{Hook:discrete_hierarchy,
    author        = "Hook, Anson",
    title         = "{Solving the Hierarchy Problem Discretely}",
    journal       = "Phys. Rev. Lett.",
    volume        = "120",
    pages         = "261802",
    year          = "2018",
    doi           = "10.1103/PhysRevLett.120.261802",
    eprint        = "1802.10093",
    archivePrefix = "arXiv",
    primaryClass  = "hep-ph"
}

@article{DiLuzio:lighter_axion,
    author        = "Di Luzio, Luca and Gavela, Bel{\'e}n
                     and Qu{\'i}lez, Pablo and Ringwald, Andreas",
    title         = "{An even lighter QCD axion}",
    journal       = "JHEP",
    volume        = "05",
    pages         = "184",
    year          = "2021",
    doi           = "10.1007/JHEP05(2021)184",
    eprint        = "2102.00012",
    archivePrefix = "arXiv",
    primaryClass  = "hep-ph"
}

@article{Peccei:1977hh,
    author  = {Peccei, R. D. and Quinn, Helen R.},
    title   = {{CP Conservation in the Presence of Pseudoparticles}},
    journal = {Phys. Rev. Lett.},
    volume  = {38},
    pages   = {1440--1443},
    year    = {1977},
    doi     = {10.1103/PhysRevLett.38.1440}
}

@article{Peccei:1977ur,
    author  = {Peccei, R. D. and Quinn, Helen R.},
    title   = {{Constraints imposed by CP conservation in the presence
                of pseudoparticles}},
    journal = {Phys. Rev. D},
    volume  = {16},
    pages   = {1791--1797},
    year    = {1977},
    doi     = {10.1103/PhysRevD.16.1791}
}

@article{Weinberg:1977ma,
    author  = {Weinberg, Steven},
    title   = {{A New Light Boson?}},
    journal = {Phys. Rev. Lett.},
    volume  = {40},
    pages   = {223--226},
    year    = {1978},
    doi     = {10.1103/PhysRevLett.40.223}
}

@article{Wilczek:1977pj,
    author  = {Wilczek, Frank},
    title   = {{Problem of Strong P and T Invariance in the
                Presence of Instantons}},
    journal = {Phys. Rev. Lett.},
    volume  = {40},
    pages   = {279--282},
    year    = {1978},
    doi     = {10.1103/PhysRevLett.40.279}
}

@article{Kim:1979if,
    author  = {Kim, Jihn E.},
    title   = {{Weak-Interaction Singlet and Strong CP Invariance}},
    journal = {Phys. Rev. Lett.},
    volume  = {43},
    pages   = {103--107},
    year    = {1979},
    doi     = {10.1103/PhysRevLett.43.103}
}

@article{Shifman:1979if,
    author  = {Shifman, Mikhail A. and Vainshtein, A. I.
               and Zakharov, Valentin I.},
    title   = {{Can Confinement Ensure Natural CP Invariance
                of Strong Interactions?}},
    journal = {Nucl. Phys. B},
    volume  = {166},
    pages   = {493--506},
    year    = {1980},
    doi     = {10.1016/0550-3213(80)90209-6}
}

@article{Dine:1981rt,
    author  = {Dine, Michael and Fischler, Willy and Srednicki, Mark},
    title   = {{A Simple Solution to the Strong CP Problem
                with a Harmless Axion}},
    journal = {Phys. Lett. B},
    volume  = {104},
    pages   = {199--202},
    year    = {1981},
    doi     = {10.1016/0370-2693(81)90590-6}
}

@article{Preskill:1982cy,
    author  = {Preskill, John and Wise, Mark B. and Wilczek, Frank},
    title   = {{Cosmology of the Invisible Axion}},
    journal = {Phys. Lett. B},
    volume  = {120},
    pages   = {127--132},
    year    = {1983},
    doi     = {10.1016/0370-2693(83)90637-8}
}

@article{Gherghetta:composite_Higgs_axion,
    author        = {Gherghetta, Tony and Nguyen, Minh D.},
    title         = {{A composite Higgs with a heavy composite axion}},
    journal       = {JHEP},
    volume        = {12},
    pages         = {094},
    year          = {2020},
    doi           = {10.1007/JHEP12(2020)094},
    eprint        = {2007.10875},
    archivePrefix = {arXiv},
    primaryClass  = {hep-ph}
}

@article{MarquesTavares:hadronic_axion,
    author        = {Marques-Tavares, Gustavo and Teo, Mae},
    title         = {{Light axions with large hadronic couplings}},
    journal       = {JHEP},
    volume        = {05},
    pages         = {180},
    year          = {2018},
    doi           = {10.1007/JHEP05(2018)180},
    eprint        = {1803.07575},
    archivePrefix = {arXiv},
    primaryClass  = {hep-ph}
}

@article{DiLuzio:astrophobic_axions,
    author        = {Di Luzio, Luca and Mescia, Federico
                     and Nardi, Enrico and Panci, Paolo
                     and Ziegler, Robert},
    title         = {{Astrophobic Axions}},
    journal       = {Phys. Rev. Lett.},
    volume        = {120},
    pages         = {261803},
    year          = {2018},
    doi           = {10.1103/PhysRevLett.120.261803},
    eprint        = {1712.04940},
    archivePrefix = {arXiv},
    primaryClass  = {hep-ph}
}

@article{Bjorkeroth:electron_decoupling,
    author        = {Bj{\"o}rkeroth, Fredrik and Di Luzio, Luca
                     and Mescia, Federico and Nardi, Enrico
                     and Panci, Paolo and Ziegler, Robert},
    title         = {{Axion-electron decoupling in nucleophobic
                      axion models}},
    journal       = {Phys. Rev. D},
    volume        = {101},
    pages         = {035027},
    year          = {2020},
    doi           = {10.1103/PhysRevD.101.035027},
    eprint        = {1907.06575},
    archivePrefix = {arXiv},
    primaryClass  = {hep-ph}
}

@article{Choi:2014rja,
    author        = {Choi, Kiwoon and Kim, Hyungjin and Yun, Seokhoon},
    title         = {{Natural inflation with multiple sub-Planckian axions}},
    journal       = {Phys. Rev. D},
    volume        = {90},
    pages         = {023545},
    year          = {2014},
    doi           = {10.1103/PhysRevD.90.023545},
    eprint        = {1404.6209},
    archivePrefix = {arXiv},
    primaryClass  = {hep-th}
}

@article{Choi:2015fiu,
    author        = {Choi, Kiwoon and Im, Sang Hui},
    title         = {{Realizing the relaxion from multiple axions and its
                      UV completion with high scale supersymmetry}},
    journal       = {JHEP},
    volume        = {01},
    pages         = {149},
    year          = {2016},
    doi           = {10.1007/JHEP01(2016)149},
    eprint        = {1511.00132},
    archivePrefix = {arXiv},
    primaryClass  = {hep-ph}
}

@article{Kaplan:2015fuy,
    author        = {Kaplan, David E. and Rattazzi, Riccardo},
    title         = {{Large field excursions and approximate discrete
                      symmetries from a clockwork axion}},
    journal       = {Phys. Rev. D},
    volume        = {93},
    pages         = {085007},
    year          = {2016},
    doi           = {10.1103/PhysRevD.93.085007},
    eprint        = {1511.01827},
    archivePrefix = {arXiv},
    primaryClass  = {hep-ph}
}

@inproceedings{bns,
    author = "Banks, Tom and Nir, Yosef and Seiberg, Nathan",
    title = "{Missing (up) mass, accidental anomalous symmetries, and the strong CP problem}",
    booktitle = "{2nd IFT Workshop on Yukawa Couplings and the Origins of Mass}",
    eprint = "hep-ph/9403203",
    archivePrefix = "arXiv",
    reportNumber = "WIS-94-14-PH, RU-94-24",
    pages = "26--41",
    month = "2",
    year = "1994"
}

@article{IAXO2019,
  author        = {Armengaud, E. and others},
  collaboration = {IAXO},
  title         = {Physics potential of the International Axion
                   Observatory ({IAXO})},
  journal       = {JCAP},
  volume        = {06},
  pages         = {047},
  year          = {2019},
  doi           = {10.1088/1475-7516/2019/06/047},
  eprint        = {1904.09155},
  archivePrefix = {arXiv},
  primaryClass  = {hep-ph}
}

@article{ORGAN2017,
  author        = {McAllister, Ben T. and others},
  title         = {The {ORGAN} Experiment: An axion haloscope above 15 {GHz}},
  journal       = {Phys. Dark Univ.},
  volume        = {18},
  pages         = {67--72},
  year          = {2017},
  doi           = {10.1016/j.dark.2017.09.010},
  eprint        = {1706.00209},
  archivePrefix = {arXiv},
  primaryClass  = {physics.ins-det}
}

@article{MADMAX2019,
  author        = {Brun, P. and others},
  collaboration = {MADMAX},
  title         = {A new experimental approach to probe {QCD} axion
                   dark matter in the mass range above $40\,\mu$eV},
  journal       = {Eur. Phys. J. C},
  volume        = {79},
  pages         = {186},
  year          = {2019},
  doi           = {10.1140/epjc/s10052-019-6683-x},
  eprint        = {1901.07401},
  archivePrefix = {arXiv},
  primaryClass  = {physics.ins-det}
}

@article{ALPHA2023,
  author        = {Millar, Alexander J. and others},
  collaboration = {ALPHA},
  title         = {Searching for dark matter with plasma haloscopes},
  journal       = {Phys. Rev. D},
  volume        = {107},
  pages         = {055013},
  year          = {2023},
  doi           = {10.1103/PhysRevD.107.055013},
  eprint        = {2210.00017},
  archivePrefix = {arXiv},
  primaryClass  = {hep-ph}
}

@article{DALI2024,
  author        = {De Miguel, Javier and Hern{\'a}ndez-Cabrera, Juan F.
                   and Hern{\'a}ndez-Su{\'a}rez, Elvio
                   and Joven-{\'A}lvarez, Enrique and Otani, Chiko
                   and Rubi{\~n}o-Mart{\'i}n, J. Alberto},
  collaboration = {DALI},
  title         = {Discovery prospects with the {Dark-photons \&
                   Axion-Like particles Interferometer}},
  journal       = {Phys. Rev. D},
  volume        = {109},
  pages         = {062002},
  year          = {2024},
  doi           = {10.1103/PhysRevD.109.062002},
  eprint        = {2303.03997},
  archivePrefix = {arXiv},
  primaryClass  = {hep-ph}
}

@article{QUAX2021,
  author        = {Alesini, D. and others},
  title         = {Search for invisible axion dark matter of mass
                   $m_a=43\,\mu$eV with the {QUAX}--$a\gamma$ experiment},
  journal       = {Phys. Rev. D},
  volume        = {103},
  pages         = {102004},
  year          = {2021},
  doi           = {10.1103/PhysRevD.103.102004},
  eprint        = {2012.09498},
  archivePrefix = {arXiv},
  primaryClass  = {hep-ex}
}

@article{ADMX2018,
  author        = {Du, N. and others},
  collaboration = {ADMX},
  title         = {Search for Invisible Axion Dark Matter with the
                   {Axion Dark Matter Experiment}},
  journal       = {Phys. Rev. Lett.},
  volume        = {120},
  pages         = {151301},
  year          = {2018},
  doi           = {10.1103/PhysRevLett.120.151301},
  eprint        = {1804.05750},
  archivePrefix = {arXiv},
  primaryClass  = {hep-ex}
}

@article{BabyIAXORADES2023,
  author        = {Ahyoune, S. and others},
  title         = {A Proposal for a Low-Frequency Axion Search in the
                   $1$--$2\,\mu$eV Range and Below with the {BabyIAXO} Magnet},
  journal       = {Annalen Phys.},
  volume        = {535},
  pages         = {2300326},
  year          = {2023},
  doi           = {10.1002/andp.202300326},
  eprint        = {2306.17243},
  archivePrefix = {arXiv},
  primaryClass  = {physics.ins-det}
}

@article{FLASH2023,
  author        = {Alesini, David and others},
  title         = {The future search for low-frequency axions and new
                   physics with the {FLASH} resonant cavity experiment
                   at {Frascati National Laboratories}},
  journal       = {Phys. Dark Univ.},
  volume        = {42},
  pages         = {101370},
  year          = {2023},
  doi           = {10.1016/j.dark.2023.101370},
  eprint        = {2309.00351},
  archivePrefix = {arXiv},
  primaryClass  = {physics.ins-det}
}

@article{DMRadio2022,
  author        = {Brouwer, L. and others},
  collaboration = {DMRadio},
  title         = {Projected sensitivity of {DMRadio}-$m^3$:
                   A search for the {QCD} axion below $1\,\mu$eV},
  journal       = {Phys. Rev. D},
  volume        = {106},
  pages         = {103008},
  year          = {2022},
  doi           = {10.1103/PhysRevD.106.103008},
  eprint        = {2204.13781},
  archivePrefix = {arXiv},
  primaryClass  = {hep-ex}
}

@article{gmanohar,
    author  = "Manohar, Aneesh and Georgi, Howard",
    title   = "{Chiral Quarks and the Nonrelativistic Quark Model}",
    journal = "Nucl. Phys. B",
    volume  = "234",
    pages   = "189--212",
    year    = "1984",
    doi     = "10.1016/0550-3213(84)90231-1"
}

@article{SRF2020,
  author        = {Berlin, Asher and D'Agnolo, Raffaele Tito
                   and Ellis, Sebastian A. R. and Nantista, Christopher
                   and Neilson, Jeffrey and Schuster, Philip
                   and Tantawi, Sami and Toro, Natalia and Zhou, Kevin},
  title         = {Axion Dark Matter Detection by Superconducting
                   Resonant Frequency Conversion},
  journal       = {JHEP},
  volume        = {07},
  pages         = {088},
  year          = {2020},
  doi           = {10.1007/JHEP07(2020)088},
  eprint        = {1912.11048},
  archivePrefix = {arXiv},
  primaryClass  = {hep-ph}
}

@misc{SRF2022,
  author        = {Berlin, Asher and others},
  title         = {Searches for New Particles, Dark Matter, and
                   Gravitational Waves with {SRF} Cavities},
  year          = {2022},
  eprint={2203.12714},
  archivePrefix={arXiv},
  primaryClass={hep-ph},
  url={https://arxiv.org/abs/2203.12714},
}

@article{WISPLC2022,
  author        = {Zhang, Zhongyue and Horns, Dieter and Ghosh, Oindrila},
  title         = {Search for dark matter with an {LC} circuit},
  journal       = {Phys. Rev. D},
  volume        = {106},
  pages         = {023003},
  year          = {2022},
  doi           = {10.1103/PhysRevD.106.023003},
  eprint        = {2111.04541},
  archivePrefix = {arXiv},
  primaryClass  = {hep-ex}
}

@misc{al,
  author       = {Ciaran O'Hare},
  title        = {cajohare/AxionLimits: AxionLimits},
  month        = jul,
  year         = 2020,
  publisher    = {Zenodo},
  version      = {v1.0},
  doi          = {10.5281/zenodo.3932430},
  howpublished = {\url{https://cajohare.github.io/AxionLimits/}}
}

@article{troy,
    author = "Cs{\'a}ki, Csaba and Roy, Tuhin S. and Ruhdorfer, Maximilian and Youn, Taewook",
    title = "{Dynamical up-quark mass generation in QCD-like theories}",
    eprint = "2505.07953",
    archivePrefix = "arXiv",
    primaryClass = "hep-ph",
    doi = "10.1103/38sb-dm8l",
    journal = "Phys. Rev. D",
    volume = "114",
    number = "1",
    pages = "016007",
    year = "2026"
}

@article{Elahi:2023anarchy,
    author        = {Elahi, Fatemeh and Elor, Gilly and Kivel, Alexey
                     and Laux, Julien and Najjari, Saereh and Yu, Felix},
    title         = {Lighter {QCD} axion from anarchy},
    journal       = {Phys. Rev. D},
    volume        = {108},
    number        = {3},
    pages         = {L031701},
    year          = {2023},
    doi           = {10.1103/PhysRevD.108.L031701},
    eprint        = {2301.08760},
    archivePrefix = {arXiv},
    primaryClass  = {hep-ph}
}

@article{Co:2024hilltop,
    author        = {Co, Raymond T. and Gherghetta, Tony
                     and Liu, Zhen and Lyu, Kun-Feng},
    title         = {A light {QCD} axion with hilltop misalignment},
    journal       = {JHEP},
    volume        = {09},
    pages         = {145},
    year          = {2024},
    doi           = {10.1007/JHEP09(2024)145},
    eprint        = {2407.12930},
    archivePrefix = {arXiv},
    primaryClass  = {hep-ph}
}

@article{Darme:selective_enhancement,
    author        = {Darm{\'e}, Luc and Di Luzio, Luca
                     and Giannotti, Maurizio and Nardi, Enrico},
    title         = {{Selective enhancement of the QCD axion couplings}},
    journal       = {Phys. Rev. D},
    volume        = {103},
    pages         = {015034},
    year          = {2021},
    doi           = {10.1103/PhysRevD.103.015034},
    eprint        = {2010.15846},
    archivePrefix = {arXiv},
    primaryClass  = {hep-ph}
}

@article{Kitano:small_instantons,
    author        = {Kitano, Ryuichiro and Yin, Wen},
    title         = {{Strong CP problem and axion dark matter
                      with small instantons}},
    journal       = {JHEP},
    volume        = {07},
    pages         = {078},
    year          = {2021},
    doi           = {10.1007/JHEP07(2021)078},
    eprint        = {2103.08598},
    archivePrefix = {arXiv},
    primaryClass  = {hep-ph}
}

@article{Redi,
    author = "Redi, Michele and Weiler, Andreas",
    title = "{Flavor and CP Invariant Composite Higgs Models}",
    eprint = "1106.6357",
    archivePrefix = "arXiv",
    primaryClass = "hep-ph",
    reportNumber = "DESY-11-109, CERN-PH-TH-154",
    doi = "10.1007/JHEP11(2011)108",
    journal = "JHEP",
    volume = "11",
    pages = "108",
    year = "2011"
}

@article{Kaplan,
    author  = "Kaplan, David B.",
    title   = "{Flavor at SSC Energies: A New Mechanism for Dynamically Generated Fermion Masses}",
    journal = "Nucl. Phys. B",
    volume  = "365",
    pages   = "259--278",
    year    = "1991",
    doi     = "10.1016/S0550-3213(05)80021-5"
}

@article{cp,
    author = "Contino, Roberto and Pomarol, Alex",
    title = "{Holography for fermions}",
    eprint = "hep-th/0406257",
    archivePrefix = "arXiv",
    reportNumber = "UAB-FT-566",
    doi = "10.1088/1126-6708/2004/11/058",
    journal = "JHEP",
    volume = "11",
    pages = "058",
    year = "2004"
}

@article{Kivel:supersizing_axions,
    author        = {Kivel, Alexey and Laux, Julien and Yu, Felix},
    title         = {{Supersizing axions with small size instantons}},
    journal       = {JHEP},
    volume        = {11},
    pages         = {088},
    year          = {2022},
    doi           = {10.1007/JHEP11(2022)088},
    eprint        = {2207.08740},
    archivePrefix = {arXiv},
    primaryClass  = {hep-ph}
}

@article{Abbott:1982af,
    author  = {Abbott, L. F. and Sikivie, P.},
    title   = {{A Cosmological Bound on the Invisible Axion}},
    journal = {Phys. Lett. B},
    volume  = {120},
    pages   = {133--136},
    year    = {1983},
    doi     = {10.1016/0370-2693(83)90638-X}
}

@article{Dine:1982ah,
    author  = {Dine, Michael and Fischler, Willy},
    title   = {{The Not-So-Harmless Axion}},
    journal = {Phys. Lett. B},
    volume  = {120},
    pages   = {137--141},
    year    = {1983},
    doi     = {10.1016/0370-2693(83)90639-1}
}

@article{Zhitnitsky:1980tq,
    author  = {Zhitnitsky, A. R.},
    title   = {{On Possible Suppression of the Axion Hadron Interactions}},
    journal = {Sov. J. Nucl. Phys.},
    volume  = {31},
    pages   = {260},
    year    = {1980},
    note    = {Original Russian version: Yad. Fiz. 31 (1980) 497--504}
}

@article{Gavela:axion_sum_rule,
    author        = "Gavela, Bel{\'e}n and Qu{\'i}lez, Pablo and Ramos, Maria",
    title         = "{The QCD axion sum rule}",
    journal       = "JHEP",
    volume        = "04",
    pages         = "056",
    year          = "2024",
    doi           = "10.1007/JHEP04(2024)056",
    eprint        = "2305.15465",
    archivePrefix = "arXiv",
    primaryClass  = "hep-ph"
}

@article{Holdom1985,
    author  = "Holdom, Bob",
    title   = "{Strong QCD at High Energies and a Heavy Axion}",
    journal = "Phys. Lett. B",
    volume  = "154",
    pages   = "316--319",
    year    = "1985",
    doi     = "10.1016/0370-2693(85)90371-5",
    note    = "Erratum: Phys. Lett. B 156 (1985) 452"
}

@article{FlynnRandall1987,
    author  = "Flynn, Jonathan M. and Randall, Lisa",
    title   = "{A Computation of the Small Instanton Contribution
                to the Axion Potential}",
    journal = "Nucl. Phys. B",
    volume  = "293",
    pages   = "731--739",
    year    = "1987",
    doi     = "10.1016/0550-3213(87)90089-7"
}

@article{Gaillard:color_unified_axion,
    author        = "Gaillard, M. K. and Gavela, M. B. and Houtz, R.
                     and Quilez, P. and del Rey, R.",
    title         = "{Color unified dynamical axion}",
    journal       = "Eur. Phys. J. C",
    volume        = "78",
    number        = "11",
    pages         = "972",
    year          = "2018",
    doi           = "10.1140/epjc/s10052-018-6396-6",
    eprint        = "1805.06465",
    archivePrefix = "arXiv",
    primaryClass  = "hep-ph"
}

@article{Csaki:many_heavy_axion,
    author        = "Cs{\'a}ki, Csaba and Ruhdorfer, Maximilian
                     and Shirman, Yuri",
    title         = "{UV Sensitivity of the Axion Mass from Instantons
                     in Partially Broken Gauge Groups}",
    journal       = "JHEP",
    volume        = "04",
    pages         = "031",
    year          = "2020",
    doi           = "10.1007/JHEP04(2020)031",
    eprint        = "1912.02197",
    archivePrefix = "arXiv",
    primaryClass  = "hep-ph"
}

@article{Gherghetta:5D_small_instantons,
    author        = "Gherghetta, Tony and Khoze, Valentin V.
                     and Pomarol, Alex and Shirman, Yuri",
    title         = "{The Axion Mass from 5D Small Instantons}",
    journal       = "JHEP",
    volume        = "03",
    pages         = "063",
    year          = "2020",
    doi           = "10.1007/JHEP03(2020)063",
    eprint        = "2001.05610",
    archivePrefix = "arXiv",
    primaryClass  = "hep-ph"
}

@article{km,
    author  = "Kaplan, David B. and Manohar, Aneesh V.",
    title   = "{Current-Mass Ratios of the Light Quarks}",
    journal = "Phys. Rev. Lett.",
    volume  = "56",
    pages   = "2004--2007",
    year    = "1986",
    doi     = "10.1103/PhysRevLett.56.2004"
}

@article{ven2,
title = {Chiral dynamics in the large N limit},
journal = {Nuclear Physics B},
volume = {171},
pages = {253-272},
year = {1980},
issn = {0550-3213},
doi = {https://doi.org/10.1016/0550-3213(80)90370-3},
url = {https://www.sciencedirect.com/science/article/pii/0550321380903703},
author = {P. {Di Vecchia} and G. Veneziano}
}

@article{ven1,
title = {U(1) without instantons},
journal = {Nuclear Physics B},
volume = {159},
number = {1},
pages = {213-224},
year = {1979},
issn = {0550-3213},
doi = {https://doi.org/10.1016/0550-3213(79)90332-8},
url = {https://www.sciencedirect.com/science/article/pii/0550321379903328},
author = {G. Veneziano}
}

@article{Csaki_Gupta_Roy,
      title={On the Dynamical Origin of the $\eta'$ Potential and the Axion Mass}, 
      author={Csaba Cs{\'a}ki and Raffaele Tito D'Agnolo and Rick S. Gupta and Eric Kuflik and Tuhin S. Roy and Maximilian Ruhdorfer},
      year={2023},
      eprint={2307.04809},
      archivePrefix={arXiv},
      primaryClass={hep-ph},
      doi = "10.1007/JHEP10(2023)139",
      journal = "JHEP",
      volume = "10",
      pages = "139",
      year = "2023" 
}

@article{RSG_Comp_Higgs,
  title = {Small instantons and the strong \textit{CP} problem in composite Higgs models},
  author = {Gupta, R. S. and Khoze, V. V. and Spannowsky, M.},
  journal = {Phys. Rev. D},
  volume = {104},
  issue = {7},
  pages = {075011},
  numpages = {10},
  year = {2021},
  month = {Oct},
  publisher = {American Physical Society},
  doi = {10.1103/PhysRevD.104.075011},
  url = {https://link.aps.org/doi/10.1103/PhysRevD.104.075011}
}

@article{QCD_axion_precisely,
   title={The QCD axion, precisely},
   volume={2016},
   ISSN={1029-8479},
   url={http://dx.doi.org/10.1007/JHEP01(2016)034},
   DOI={10.1007/jhep01(2016)034},
   number={1},
   journal={Journal of High Energy Physics},
   publisher={Springer Science and Business Media LLC},
   author={di Cortona, Giovanni Grilli and Hardy, Edward and Vega, Javier Pardo and Villadoro, Giovanni},
   year={2016},
   month=jan }

@article{AGASHE2005165,
title = {The minimal composite Higgs model},
journal = {Nuclear Physics B},
volume = {719},
number = {1},
pages = {165-187},
year = {2005},
issn = {0550-3213},
doi = {https://doi.org/10.1016/j.nuclphysb.2005.04.035},
url = {https://www.sciencedirect.com/science/article/pii/S0550321305003445},
author = {Kaustubh Agashe and Roberto Contino and Alex Pomarol},
}

@article{Ferretti_2014,
   title={UV completions of partial compositeness: the case for a SU(4) gauge group},
   volume={2014},
   ISSN={1029-8479},
   url={http://dx.doi.org/10.1007/JHEP06(2014)142},
   DOI={10.1007/jhep06(2014)142},
   number={6},
   journal={Journal of High Energy Physics},
   publisher={Springer Science and Business Media LLC},
   author={Ferretti, Gabriele},
   year={2014},
   month=jun }

@book{Panico_2016,
   title={The Composite Nambu-Goldstone Higgs},
   ISBN={9783319226170},
   ISSN={1616-6361},
   url={http://dx.doi.org/10.1007/978-3-319-22617-0},
   DOI={10.1007/978-3-319-22617-0},
   journal={Lecture Notes in Physics},
   publisher={Springer International Publishing},
   author={Panico, Giuliano and Wulzer, Andrea},
   year={2016} }

@article{Harlow:2018jwu,
    author = "Harlow, Daniel and Ooguri, Hirosi",
    title = "{Constraints on Symmetry from Holography}",
    eprint = "1810.05337",
    archivePrefix = "arXiv",
    primaryClass = "hep-th",
    doi = "10.1103/PhysRevLett.122.191601",
    journal = "Phys. Rev. Lett.",
    volume = "122",
    pages = "191601",
    year = "2019"
}

@misc{antusch2026,
      title={Updated Running Quark and Lepton Parameters at Various Scales}, 
      author={Stefan Antusch and Kevin Hinze and Shaikh Saad},
      year={2026},
      eprint={2510.01312},
      archivePrefix={arXiv},
      primaryClass={hep-ph},
      url={https://arxiv.org/abs/2510.01312}, 
}

@article{tHooft_PRD,
  title = {Computation of the quantum effects due to a four-dimensional pseudoparticle},
  author = {'t Hooft, G.},
  journal = {Phys. Rev. D},
  volume = {14},
  issue = {12},
  pages = {3432--3450},
  numpages = {0},
  year = {1976},
  month = {Dec},
  publisher = {American Physical Society},
  doi = {10.1103/PhysRevD.14.3432},
  url = {https://link.aps.org/doi/10.1103/PhysRevD.14.3432}
}
\end{document}